\documentclass[a4paper]{aa}
\usepackage{graphicx}
\usepackage{natbib}
\def\aap{A\&A}

\def\apj{ApJ}

\def\apss{ApSS}

\begin{document}
   \title{First VLTI/MIDI observations of a Be star: $\alpha$ Ara \thanks{Based on observations
       made with the Very Large Telescope Interferometer at
       Paranal Observatory}}

\titlerunning{$\alpha$ Ara observed by MIDI}

   \author{O.~Chesneau \inst{1},\inst{2}
            \and
          A.~Meilland \inst{2}
            \and
            T.~Rivinius \inst{3}
            \and
          Ph.~Stee \inst{2}
        \and
          S.~Jankov \inst{4}
            \and
          A.~Domiciano~de~Souza \inst{5}
            \and
         U.~Graser \inst{1}
            \and
           T.~Herbst \inst{1}
            \and
          E.~Janot-Pacheco \inst{6}
            \and
          R.~Koehler \inst{1}
            \and
        C.~Leinert \inst{1}
            \and
    S.~Morel \inst{7}
            \and
    F.~Paresce \inst{7}
            \and
     A.~Richichi \inst{7}
        \and
          S.~Robbe-Dubois \inst{4}
          }

   \offprints{O. Chesneau}

\institute{Max-Planck-Institut f\"{u}r Astronomie, K\"{o}nigstuhl
17, D-69117 Heidelberg, Germany\\
\email{Olivier.Chesneau@obs-azur.fr}
    \and
Observatoire de la C\^{o}te d'Azur-CNRS-UMR 6203 Avenue Copernic,
Grasse, France
    \and
    Landessternwarte Heidelberg, K\"{o}nigstuhl 12, 69117 Heidelberg,
    Germany
    \and
Laboratoire Universitaire d'Astrophysique de Nice, France
    \and
Max-Planck-Institut f\"{u}r Radioastronomie, Auf dem H\"{u}gel 69,
53121 Bonn, Germany
    \and
Instituto de Astronomia, Geofi\'sica e Ci\^{e}ncias Atmosf\'{e}ricas
da Universidade de S\~{a}o Paulo (IAG-USP),\\
CP 9638,01065-970 S\~{a}o Paulo, Brazil \and European Southern
Observatory, Karl-Schwarzschild-Strasse 2, D-85748 Garching,
Germany }

   \date{Received; accepted }

   \abstract{We present the first VLTI/MIDI observations of the Be
   star $\alpha$~Ara, showing a nearly unresolved circumstellar disk
   in the N band. The interferometric measurements made use of the UT1
   and UT3 telescopes. The projected baselines were 102 and
   74 meters with position angles of 7$\degr$ and 55$\degr$,
   respectively. These measurements put an upper limit to the envelope
   size in the N band under the Uniform disk approximation of $\phi_{\rm max}= 4\pm1.5$\,mas, corresponding to
   14\,$R_{\star}$, assuming $R_{\star}$=4.8${\rm R}_\odot$ and the
   Hipparcos distance of 74\,pc.

   On the other hand the disk density must be large enough to produce
   the observed strong Balmer line emission. In order to estimate the possible circumstellar and stellar
   parameters we have used the SIMECA code developed by Stee
   \cite{Stee1} and Stee \& Bittar \cite{Stee3}. Optical spectra
   taken with the \'{e}chelle instrument {\sc Heros} and the ESO-50cm
   telescope, as well as infrared ones from the 1.6m Brazilian
   telescope have been used together with the MIDI spectra and
   visibilities. These observations put complementary constraints on
   the density and geometry of $\alpha$ Ara circumstellar disk. We discuss on
   the potential truncation of the disk by a companion and we present
   spectroscopic indications of a periodic perturbation of some Balmer lines.

   }

   \maketitle
%

\section{Introduction}
Be stars are hot stars exhibiting the so-called Be phenomenon,
i.e. Balmer lines in emission and infrared excess.
Optical/infrared and ultraviolet observations of Be stars have
been widely interpreted as evidence for two quite distinct regions
in the circumstellar environment of these objects: a rotating,
dense equatorial region, called "disk" in the following, and a
diluted polar region which expands with velocities that may reach
2000 ${\rm km\,s}^{-1}$ (e.g. Marlborough \citealp{marlborough}).

The thermal infrared domain is an important spectral region where
the transition between optically thin and thick disk occurs. The
optically thick disk emission is growing from 7 to 15\,$\mu$m. It
reaches about half of the total continuum flux  at 8~$\mu$m and
dominates the Spectral Energy Distribution (SED) almost completely
at 15-20\,$\mu$m (Gehrz~et~al.~\citealp{gehrz}). The flux from the
central star at these wavelengths is 10 to 50 times fainter
compared to $\lambda = 2\,\mu{\rm m}$ whereas  the ratio $F_{\rm
env}/F_\star$ reaches a factor 4-10. In particular for nearly
edge-on disks, the flux from the optically thick disk is very
sensitive to the inclination angle, since it is proportional to
the emitting surface.

Observations at 20~$\mu$m by Gehrz~et~al.~\cite{gehrz}
demonstrated that the mid-IR excess is due to free-free emission
from the hydrogen envelope.  The ad-hoc model from Waters
\cite{Waters86} has been successful to explain near- and far-IR
observations and is coherent with polarization data (Cot\'{e} \&
Waters \citealp{Cote87}, Waters \& Marlborough \citealp{waters92},
Dougherty et al. \citealp{dougherty}, Cot\'{e} et
al.~\citealp{Cote96}). In this model the IR excess originates from
a disk with an opening angle $\theta$ and a density distribution
$\rho(r) \propto r^{-n}$ seen at an observer's angle $i$. From
IRAS data of four Be stars Waters found that the far-IR slope of
the SED is not strongly influenced by the opening angle, as long
as $\theta + i \le 90^\circ$. Moreover, he found that n=2.4 gives
a good agreement for $\chi$ Oph, $\delta$ Cen and $\phi$ Per if
the disk is truncated to 6.5 R$_{*}$.



Interferometry in the visible has also been used to study the
circumstellar environment of Be stars (Thom et al.~\citealp{thom},
Mourard et al. \citealp{mourard}, Quirrenbach et
al.~\citealp{quirrenbach1}; Stee et al. \citealp{Stee1}). Quirrenbach
et al. \cite{quirrenbach2} gave an upper limit on the opening angle of
about 20$\degr$ and using spectral Differential Interferometry (DI)
structures within Be disks were monitored with a high spatial resolution
during several years (Vakili et al. \citealp{Vakili98}, Berio et al. \citealp{berio}).
Nevertheless, the disk extension at IR wavelengths is still subject to an active
debate since various authors give quite different extensions as a
function of wavelength. For instance Waters \cite{Waters86} found a
typical IR extension of $16\,R_{\star}$ from 12, 25 and 60\,$\mu$m
observations, Gehrz et al. \cite{gehrz} found $8\,R_{\star}$ at 2.3
and 19.5\,$\mu$m, Dougherty et al. \cite{dougherty} a disk size larger
than $20\,R_{\star}$ in the near-IR, and finally Stee \& Bittar
\cite{Stee3} obtained that the Br$\gamma$ emission line and the nearby
continuum originates from a very extended region with a typical size
of $40\,R_{\star}$.

In order to solve this issue, we have used the VLTI/MIDI
interferometer to obtain the first IR measurements in the N
spectral band of a Be star.  The selected target,
\object{$\alpha$\,Ara} (HD\,158\,427, HR\,6510, B3\,Ve) is one of
the closest Be stars with an estimated distance of $74\pm6$\,pc
derived from the Hipparcos parallax, and is well known as emission
line star since the discovery of its H\,$\beta$ emission by
Pickering (\citealp{pick96}, \citealp{pick96}). $\alpha$\,Ara is a
2.9 magnitude star in the Johnson V-Band and its IRAS flux at 12
micron is 12.7~Jy. Its color excesses $E(V-L)$ and
$E(V-12\,\mu{\rm m})$ are respectively 1.8 and 2.23, being among
the highest of its class. The projected rotational velocity $v
\sin i$ has been estimated to be 250--300\,km\,s$^{-1}$ (Yudin
\citealp{yudin2}, Chauville et al. \citealp{chauville}). An
intrinsic linear polarization of Pl$\approx$0.6\% with a Position
Angle (PA) of 172$^\circ$ has also been detected (McLean \& Clarke
\citealp{mclean}, Yudin \citealp{yudin2}). Since the disk
orientation is expected to be perpendicular to this direction, we
expect the disk major-axis to be around PA$\approx82^\circ$ (Wood
et al.  \citealp{wood96a}, \citealp{wood96a}, Quirrenbach et al.
\citealp{quirrenbach2}).

The spectral type definition of $\alpha$\,Ara is constant in the
literature oscillating between B2\,Ve and B3\,Ve. We adopted
B3\,Ve following the latest study of Chauville et al.
\cite{chauville}. The typical stellar radius and effective
temperature for this spectral type are 4.8~R$_{\sun}$ and $T_{\rm
eff}=18\,000$ respectively.

By using IRAS observations of $\alpha$\,Ara, Waters
\cite{Waters86} estimated the outer disk radius is estimated to be
about 7~R$_{\star}$, i.e. 2.1\,mas at a distance of 74~pc. Stee
\cite{Stee4} predicts the visibility of $\alpha$~Ara at 2~$\mu$m
to be lower than 0.2 with a 60m baseline, i.e. fully resolved.
Using the same model parameters, a significant visibility loss
should still be detectable in the N band. This paper will show
that at 10 $\mu$m $\alpha$~Ara is in fact unresolved which,
combined with spectroscopic data, puts some constraints on
density, inclination angle, disk orientation and distance of
$\alpha$~Ara.

The paper is organized as follows. In Section \ref{secobs} we present
the interferometric MIDI observations and the H-band spectroscopy.
Section \ref{secsimeca} briefly describes the SIMECA code and the
envelope parameters used for the modeling of $\alpha$~Ara.
The modeling strategy is exposed in details. In Section \ref{secresult} we
present our theoretical results and discuss the possibility for a close but unseen companion
to truncate the disk making $\alpha$~Ara unresolved in the N
band. Section \ref{sec:discussion} presents some spectroscopic
variations of the H$\beta$ line profile that may be another evidence for
this hypothetical companion. Finally, Section \ref{secconcl} draws the
main conclusions from these first IR interferometric measurements of a Be
star.

\section{Observations}
\label{secobs}
\subsection{Interferometric data}
We use the VLTI/MIDI Interferometer described by
Leinert et al. (\citealp{2003Msngr.112...13L},
\citealp{2003Ap&SS.286...73L}) which combine the mid-IR light (N band) from
the two VLT Unit Telescopes Antu (UT1) and Melipal (UT3).
The observations of $\alpha$~Ara were done
during the nights of June, 16th and 17th 2003 and
are presented in Table~\ref{tabobs}.

\begin{table}[bt]
 \caption{\label{tabobs}Journal of observations.}
\vspace{0.3cm}
\begin{center}
 \begin{tabular}{l|l|c|r}
 \hline
  \hline
\multicolumn{4}{c}{June, 16 2003 night, B=102m, PA=7$\degr$}\\
 \hline
 Star & Template & Time &Frames \\
 \hline
HD\,165\,135&tracking&00:36:11& 15000 \\
HD\,165\,135&phot. UT1&00:42:33& 2000 \\
HD\,165\,135&phot. UT3&00:44:12& 2000\\
HD\,152\,786&tracking&01:06:42& 12000\\
HD\,152\,786&phot. UT1&01:12:15& 3000\\
HD\,152\,786&phot. UT3&01:14:33& 3000\\
$\alpha$\,Ara&tracking&01:35:27&12000\\
$\alpha$\,Ara&phot. UT1&01:40:55&3000 \\
$\alpha$\,Ara&phot. UT3&01:43:12& 3000\\
HD\,165\,135&tracking&02:46:05&12000 \\
HD\,165\,135&phot. UT1&02:51:32&3000 \\
HD\,165\,135&phot. UT3&02:46:05&3000 \\
$\alpha$\,Aql&tracking&06:44:35&12000 \\
$\alpha$\,Aql&phot. UT1&06:50:08&3000 \\
$\alpha$\,Aql&phot. UT3&06:52:14&3000 \\
 \hline
\multicolumn{4}{c}{June, 17 2003 night, B=79m, PA=55$\degr$}\\
 \hline
HD\,139\,997&tracking&05:31:17& 10000 \\
HD\,139\,997&phot. UT1&05:35:38& 2000 \\
HD\,139\,997&phot. UT3&05:37:29& 2000\\
HD\,168\,545&tracking&06:35:05& 15000 \\
HD\,168\,545&phot. UT1&06:40:33& 2000 \\
HD\,168\,545&phot. UT3&06:42:21& 2000\\
$\alpha$\,Ara&tracking&07:09:19&12000\\
$\alpha$\,Ara&phot. UT1&07:14:47&3000 \\
$\alpha$\,Ara&phot. UT3&07:17:05& 3000\\
\hline
  \end{tabular}
\end{center}
 \end{table}

On the 16th of June, $\alpha$~Ara was observed with a
102~m baseline and a PA=7$\degr$ whereas June, 17 the projected
baseline was 79~m and PA=55$\degr$.

\begin{table}[b!]
\begin{center}
\caption{Calibrator stars parameters, from the
MIDI calibrators catalogue, (Van Boekel et al. 2005).}
\label{tab:calibrators}
\begin{tabular}{lcrc}

\hline \hline

Calibrator & Spectral &  Flux$^i$ & Uniform disk\\
           &  type    &   (Jy)& diameter (mas)\\
\hline
HD\,139\,997 &  K5\,III &  14.5 &3.46~$\pm$~0.38\\
HD\,152\,786 &  K3\,III &  82.2&  7.21~$\pm$~0.21\\
HD\,165\,135 &  K0\,III & 16.3 &3.33~$\pm$~0.05\\
HD\,168\,454 &  K3\,IIIa&43.7 & 5.78~$\pm$~0.15\\
\hline
\end{tabular}
\end{center}\vskip-3mm

{\scriptsize $^{i}$ IRAS 12.5 $\mu$m flux}

\end{table}

The observing sequence, described extensively in Przygodda et al.
\cite{2003Ap&SS.286...85P} is summarized hereafter. The chopping
mode (f=2Hz, angle $-90\degr$) is used to point and visualize the
star. The number of frames recorded by image are generally 2000,
the exposure time is by default 4\,ms in order to avoid the quick
background saturation. If the result of the acquisition sequence
is not satisfactory, this procedure is re-executed. Then, the MIDI
beam combiner, the wide slit (0$\farcs6 \times$2$\arcsec$), and
the NaCl prism are inserted to disperse the light and search for
the fringes by moving the VLTI delay lines. The resulting spectra
have a resolution $\lambda$/$\Delta \lambda \sim$30.

When searching for the fringe signal, the delay line of the VLTI
is moved, while the MIDI internal piezo-driven delay line is
performing additional scans of the fringe pattern. Once the
fringes are found another file is recorded while MIDI is tracking
the fringes by its own, i.e. by performing a real time estimation
of the OPD based on the data continuously recorded. The correction
is send to the VLTI delay lines at a rate of about 1Hz. Finally,
the photometric data are recorded in two more files.

In the first file, only one shutter is opened, corresponding to the
calibration of the flux from the first telescope, here UT1, and the
flux is then divided by the MIDI beam splitter and falls on two
different regions of the detector. The total flux is determined
separately by chopping between the object and an empty region of the
sky, then the source flux is computed by subtraction.  Once it has been
done, the same procedure is carried out again for the beam arriving
from UT3.

We performed the spectral reduction and fringe calibration using a
software developed at the Max-Planck Institut f\"{u}r Astronomie
in Heidelberg, written in IDL\footnote{Package available at this
address: http://www.mpia-hd.mpg.de/MIDISOFT/.}. The reader is
referred to Leinert et al. \cite{leinert} for an extensive
discussion on MIDI data reduction and error handling.

\begin{figure}
  \begin{center}
      \includegraphics[height=7.cm]{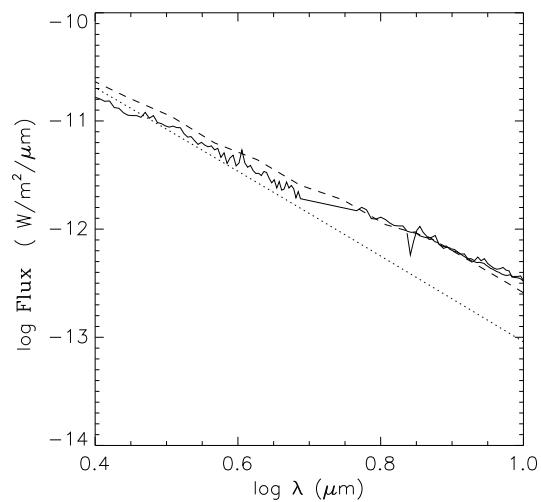}
  \end{center}
  \caption{$\alpha$ Ara flux measured by the VLTI/MIDI (8-13 $\mu$m)
  instrument with ISO data (2.5-11.5 $\mu$m) plotted as solid line
  compared with the theoretical SED of the star alone (dashed line)
  and the star+envelope emission (dotted line) assuming a distance of
  105\,pc.  Note the IR excess starting around 5 $\mu$m, i.e. log
  $\lambda$=0.7.}
\label{SED}
\end{figure}

\begin{figure*}
  \begin{center}
      \includegraphics[height=7.cm]{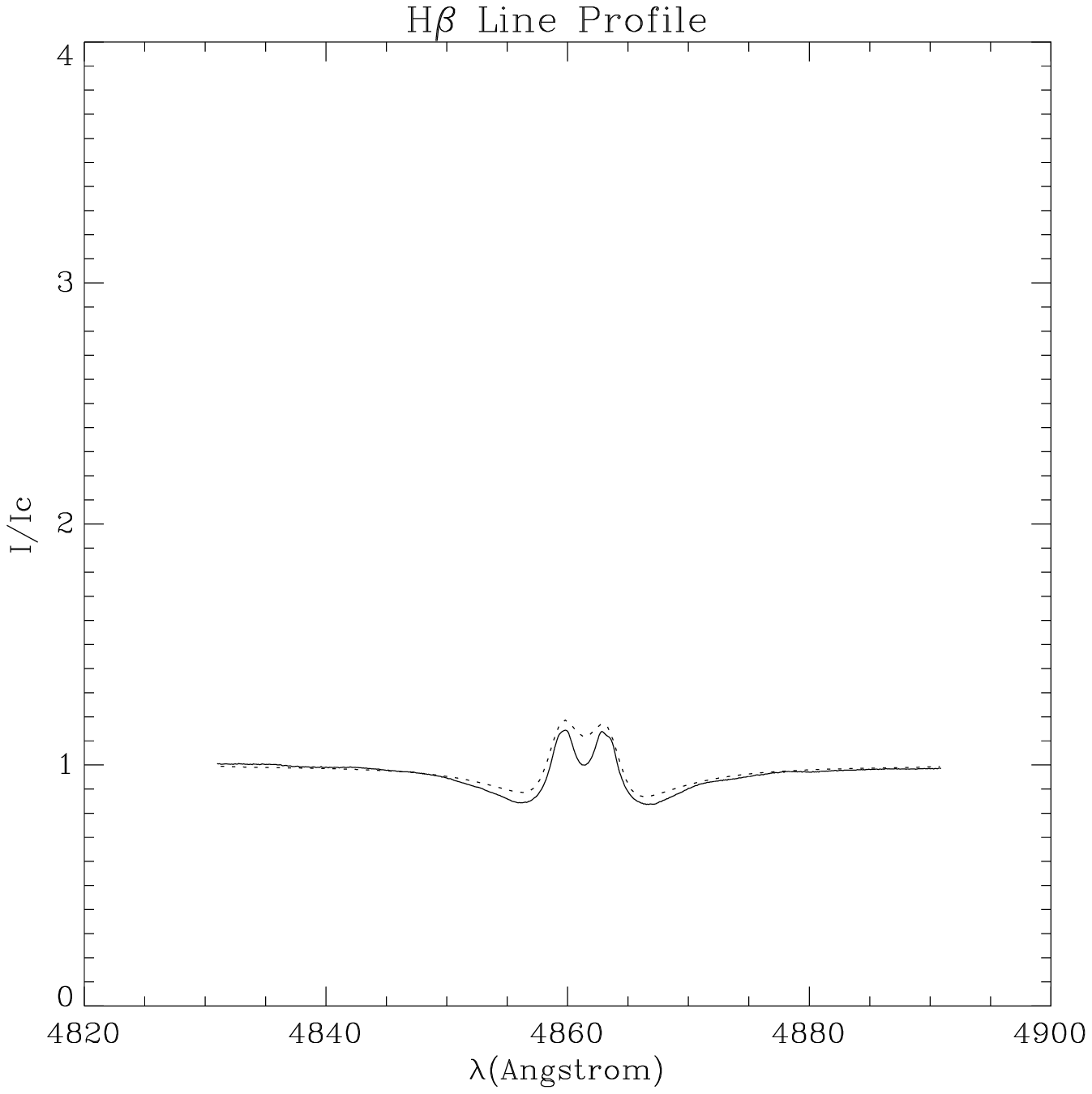}
      \includegraphics[height=7.cm]{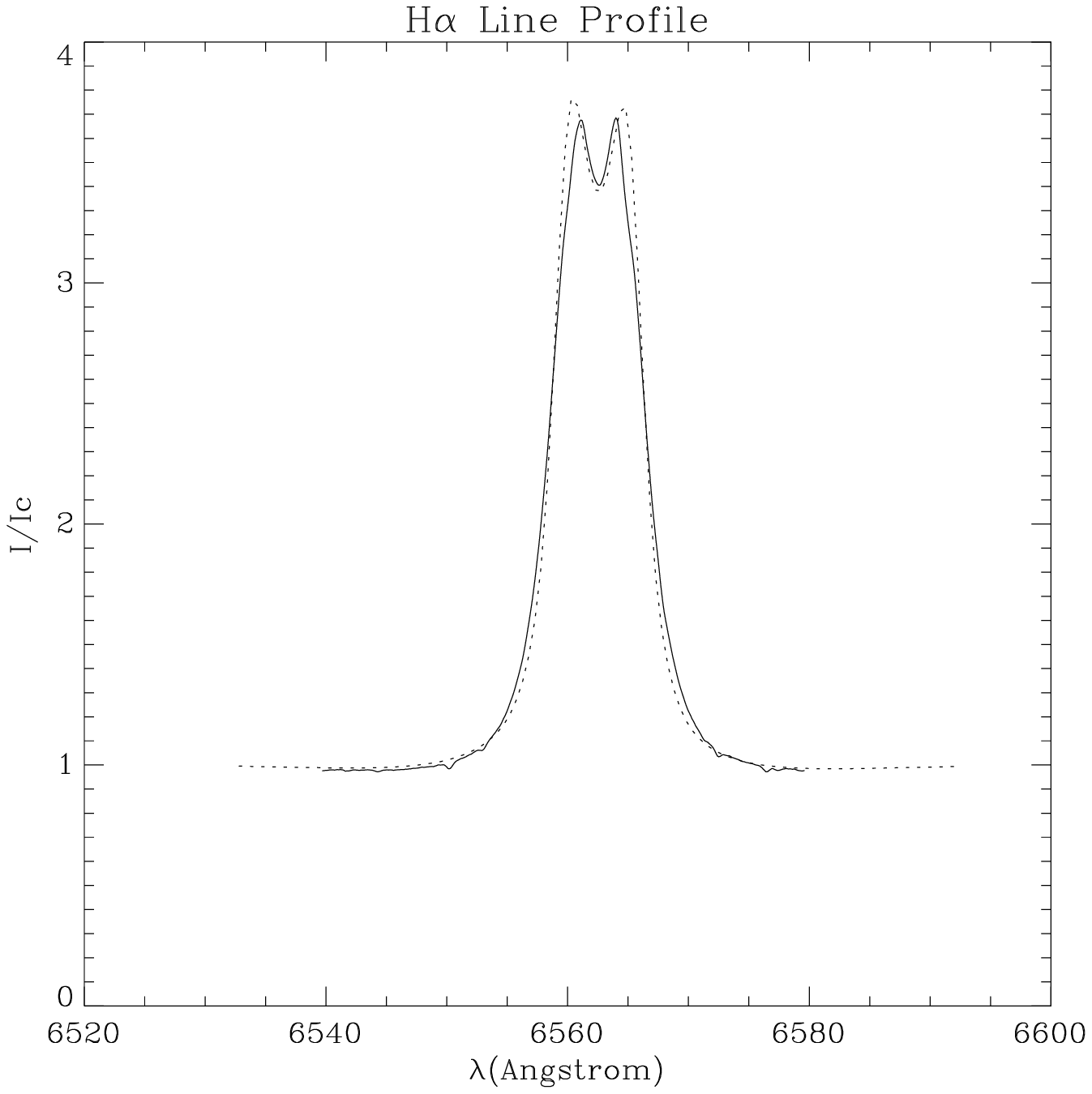}
  \end{center}
  \caption{H$\alpha$ and H$\beta$ line profiles observed in mid-1999
  (solid lines) compared with the best SIMECA fits (dotted lines).}
\label{halpha}
\end{figure*}

The photometric datasets used for the calibration of the contrast
of the dispersed fringes are also used for the spectrum flux
calibration. The frames obtained on the target and the ones
obtained on the sky are averaged. Then, the averaged sky frame is
subtracted from the averaged target frame. In the resulting
integration, the spectral axis is oriented along the horizontal
detector axis. For each detector column, a Gaussian function is
fitted to the peak. The position of the spectrum in all
illuminated columns, as function of the column number, is fitted
by a quadratic polynomial. In a similar way, the width of the
spectrum is fitted by a linear function. This procedure is carried
out on both photometric datasets, corresponding to telescopes UT1
and UT3, respectively. Both fits are then averaged and used to
create a mask consisting of a Gaussian function with the average
position and width of the spectra for each column. Finally, a
weighting function is computed to create a mask for the
extraction.

The photometric masks are also used to extract the fringes
information. Each frame of the fringe data is reduced to a
one-dimensional spectrum by multiplying it with the mask;
integrating in the direction perpendicular to the spectral
dispersion; and finally subtracting the two output channels of the
beam combiner from each other. The spectra from each scan with the
piezo-mounted mirrors are collected and Fourier-transformed from
optical path difference (OPD) to frequency space. The fringe
amplitude at each wavelength is then obtained from the power
spectrum. We typically co-added the signal by bin of four pixels
in the dispersion direction to improve the signal-to-noise ratio.

The scans where fringes were actually detected are selected based on
the white-light fringe amplitude, i.e.\ the fringe amplitude that is
seen after integrating over all usable wavelengths. The histogram of
all white-light fringe amplitudes within a fringe track dataset
usually shows a small peak near zero, and a broad peak at higher
amplitudes. We interactively set a threshold just below this broad
peak, and average the fringe amplitudes of all scans with a
white-light fringe amplitude higher than this threshold to give the
final fringe amplitude as a function of wavelength. The raw visibility
is obtained by dividing the fringe amplitude by the total flux. The
calibrated visibility is obtained by dividing the raw visibility of an
object by that of a calibrator. The photometrically calibrated flux
creating the fringes is called correlated flux.

The interferometric and photometric data are presented in Table~1
and the parameters of the interferometric calibrators are given
in Table~2.

The flux calibration has been performed with the data of June, 16
only. $\alpha$~Aql was considered as a good calibrator for the
absolute flux extraction and we have used the ISO observations of
this star as a flux reference (but has not been used as
interferometric calibrator). An independent calibration is
performed for the individual spectra from each part of the
detector and each telescopes. The airmass is also corrected
independently for each of them by using the observations of one
calibrator for two different times. Fig.~\ref{SED} shows the
VLTI/MIDI Spectral Energy Distribution (SED) with the ISO data and
the theoretical SED obtained with the SIMECA code (see hereafter).

The errors on the observed visibilities are mostly systematic. The
statistical signal-to-noise ratio on the white light fringe
amplitudes is 5-10 at minimum and much better after adding up the
several hundred scans taken per interferometric measurement.
Most of the time, the dominant error source is the fluctuations of the
photometry which occur between the fringe recording and the
photometry recording from the individual telescopes which
are separated in time by 2 to 10 minutes. This affect the fringe
signal for object and calibrator measurements. Comparing the raw
visibilities observed for different calibrator stars during one
night, the standard deviation of these values under good
conditions amounts to $\pm$~5-10~\% (relative) at the red and blue
end of the spectrum while for adverse conditions
these numbers have to be multiplied by a factor of 2-2.5. These
fluctuations are much larger for instance than the systematic
error expected from the diameters of the calibrators
and are mostly achromatic (Leinert et al. \citealp{leinert}). Yet,
the instrumental visibility curve of MIDI is very stable in shape,
i.e. the slope of the curve is typically varying by less than
3-5\%. The visibilities observed on calibrators (``instrumental
visibility'') rises from about 0.4 at 8~$\mu$m to about 0.7 at
13~$\mu$m, rather repeatable from night to night (i.e. it means
that, in good atmospheric conditions, the absolute level is dependant on
the photometric variations and hence the stability of the
atmospheric conditions but the shape of the visibility curves is
mostly an instrumental parameter only weakly affected by the
atmosphere).

The data reduction described previously does not consider the
variations of the photometry during the 1-2min needed to record
the data and the visibility estimation is usually based on the
mean photometric flux. Moreover, under good atmospheric
conditions, the statistics of the correlated flux per individual
scan follow roughly a Gaussian curve and the mean of this
correlated flux is used for the visibility estimation. The main
origin of the variability of the correlated flux is the high
frequency photometric fluctuations, the atmospheric piston and the
quality of the fringe tracking (i.e. the instrumental piston).

During the night of June, 17 2003  thin cirrus were passing
through the telescope beams. The seeing was quite stable, below
0.5$\arcsec$ during all the night but the standard deviation of
the flux from the target pointed by the (visible) seeing
monitor\footnote{This information has been extracted from the ESO
Ambient conditions database of Paranal observatory:
http://archive.eso.org/.} was oscillating between 0.02 and 0.05
between 3h and 8h (UT time), indicating the passage of thin
cirrus. These fluctuations seem correlated with an acceleration of
the atmosphere turbulence with a coherence time in visible
decreasing from $\tau_0$=5.5~ms at 3h UT to $\tau_0$=2~ms at 7h30
UT. By comparison, the photometric fluctuations of the seeing
monitor calibrator during the 16th of June are below 0.01 during
the entire night although the turbulence was also quite rapid with
$\tau_0$=3~ms. The variability of the atmosphere during the June,
17 night was such that a careful inspection of the stability of
the photometry and the correlated flux recorded has been
necessary. The passages of the thin cirrus were easily recognized
by a rapid drop of the flux lasting between 0.5 and 2s during the
photometric record and by a simultaneous photometric and
correlated fluxes drop in the tracking files. The frames and the
scans concerned were excluded from the data reduction process:
this represent about 60\% of the photometry and 50\% of the scans,
i.e. about 800 frames were left from each photometric files and
about 100 scans were used for the visibility estimation. Of
course, this treatment, applied to $\alpha$~Ara and its
calibrators is a first order correction and the standard deviation
of the instrumental visibility curves of the calibrators observed
between 2h UT and 9h UT is still large compared to good observing
night, on the order of 18\% whereas the standard deviation of the
previous night is only 8\%. Fortunately, the longest baseline of
102m has been used during the best night. The error bars reported
in Fig.~\ref{visi_SIMECA} and Fig.~\ref{visi_TRUNC} reflect the
standard deviation computed by this mean.

\subsection{Spectroscopic data}
For an investigation of the variability of $\alpha$\,Ara this
study makes use of a set of 42 echelle spectra, taken during 69
nights in May to July 1999. The data were secured with the {\sc
Heros} instrument, attached to the ESO-50cm telescope. The
resolving power is $\Delta\lambda / \lambda = 20\,000$, recording
the regions from 350 to 565\,nm and from 585 to 860\,nm in a blue
and red channel simultaneously. The spectra have a typical
signal-to-noise ratio of 170 (Fig.\ref{halpha}).

Next to these data, complementary spectra have also been recorded
quasi-simultaneously with the VLTI run in order to know whether
$\alpha$ Ara had shown IR emission lines. We have observed
$\alpha$~Ara in the J2 band (1.2283-1.2937$\mu$m) using the 1.6 m
Perkin-Elmer telescope and Coud\'e spectrograph(with R=10\,000) at
the Observat\'orio do Pico dos Dias, Laborat\'orio Nacional de
Astrof\'{\i}sica (LNA), Itajub\'a, Brasil. In total, five stellar
spectra were recorded with the C\^amara Infravermelho (CamIV)
detector, August 13, 2003. Each spectrum corresponds to a
different position of the star along the spatial axis of the
entrance slit, allowing to use the median of the five
two-dimensional frames, from which the average bias frame was
subtracted, as a sky background. After dividing the stellar and
sky background frames with the average flat field, the sky was
subtracted from two-dimensional stellar spectra. The extraction of
one-dimensional spectra, as well as wavelength calibration, using
an Ar-Ne calibration lamp, have been performed with standard
IRAF\footnote{IRAF is distributed by the National Optical
Astronomy Observatories, which is operated by the Association of
Universities for Research in Astronomy (AURA), Inc., under
cooperative agreement with the National Science Foundation.}
packages. Finally, the continuum was normalized to unity.

The observed Pa$\beta$ line profile is shown in
Figure~\ref{pbeta_bad}. The emission intensity was $I/I_{\rm
c}$~$\approx$ 2.3, while the violet-to-red peak height ratio $V/R$
was 1.17. Even if the spectroscopic data were not recorded fully
simultaneously with the interferometric ones, available
observations of typical timescales of the spectral changes of
$\alpha$~Ara, such as the above described {\sc Heros} data,
justify the assumption that the circumstellar environment of
$\alpha$~Ara has not changed significantly between June and August
2003. This Pa$\beta$ line profile, combined with the
interferometric data, constrains the physical parameters of the
circumstellar environment of $\alpha$~Ara, as it will be shown in
the following sections.

\begin{table}
\caption{Stellar parameters used for the modelling of $\alpha$ Ara
with SIMECA\label{parameters}} {\centering \begin{tabular}{cc}
\hline
Spectral type& B3 Ve\\
$T_{\rm eff}$& 18\,000\,K\\
Mass& 9.6 M\( _{\sun } \)\\
Radius& 4.8 R\( _{\sun } \)\\
Luminosity& 2.2 10\( ^{3} \)L\( _{\sun } \)\\
\hline
\end{tabular}\par}
\end{table}

\section{The SIMECA code and the envelope parameters used for the
  modeling of $\alpha$ Ara}
\label{secsimeca} In order to constrain the physical parameters of
the circumstellar environment of $\alpha$ Ara, we have used the
SIMECA code. This code, described in previous paper (see Stee \&
Ara\`ujo \citealp{Stee0}; Stee et al. \citealp{Stee1}; Stee \&
Bittar \citealp{Stee3}), has been developped to model the
environment of active hot stars.  SIMECA computes line profiles,
Spectral Energy Distributions (SEDs) and intensity maps, which can
directly be compared to high angular resolution observations.  The
envelope is supposed to be axi-symmetric with respect to the
rotational axis.  No meridional circulation is allowed. We also
assume that the physics of the polar regions is well represented
by a CAK type stellar wind model (Castor et al. \citealp{Castor})
and the solutions for all stellar latitudes are obtained by
introducing a parameterized model which is constrained by the
spectroscopic and interferometric data. The inner equatorial
region is dominated by rotation, therefore being quasi Keplerian.
The ionization-excitation equations are solved for an envelope
modeled in a 410{*}90{*}71 cube. In order to reproduce the
Pa$\beta$ profile we have computed the line corresponding to the
transitions between the 5-3 atomic levels. The populations of the
atomic levels are strongly altered from their Local Thermodynamic
Equilibrium (LTE) values by non-LTE conditions in the wind. For
computation, we start with the LTE populations for each level, and
then compute the escape probability of each transition in the
wind, obtaining up-dated populations. By using these populations
as input values for the next step, we iterate until convergence.
The basic equations of the SIMECA code are given in detail by Stee
et al. \cite{Stee1}.

To take into account the photospheric absorption line, we assume
the underlying star to be a normal B3 Ve star with $T_{\rm eff} =
18\,000$\,K and $R=4.8 {\rm R}_{\odot}$ and synthesize the
photospheric line profiles using the SYNSPEC code by Hubeny
(Hubeny \citealp{Hubeny1}; Hubeny \& Lanz \citealp{Hubeny2}). The
resulting line profile is broadened by solid-body rotation and
might be further altered by absorption in the part of the envelope
in the line of sight towards the stellar disk.

\begin{figure*}
  \begin{center}
      \includegraphics[height=7.cm]{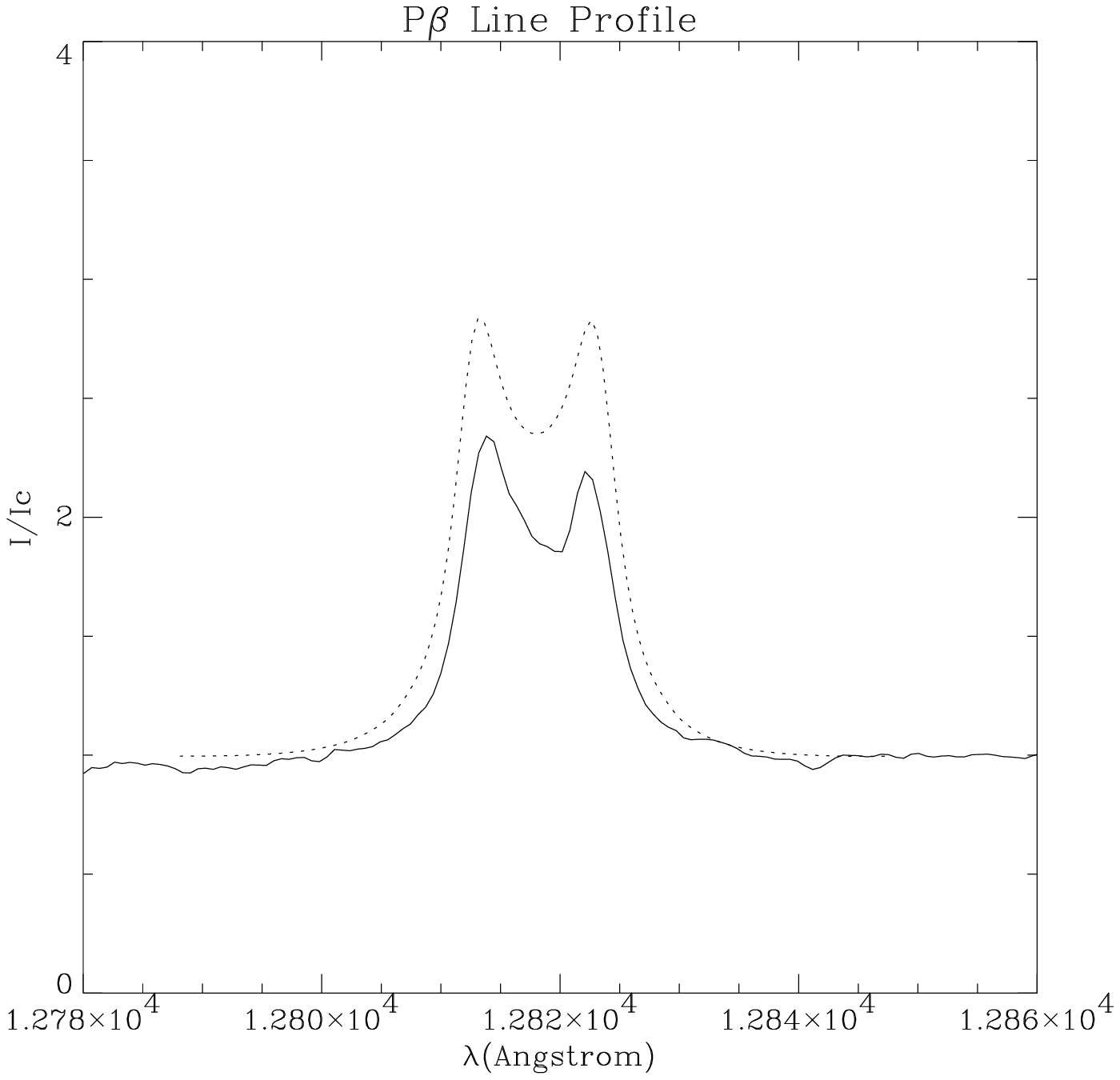}
      \includegraphics[height=7.cm]{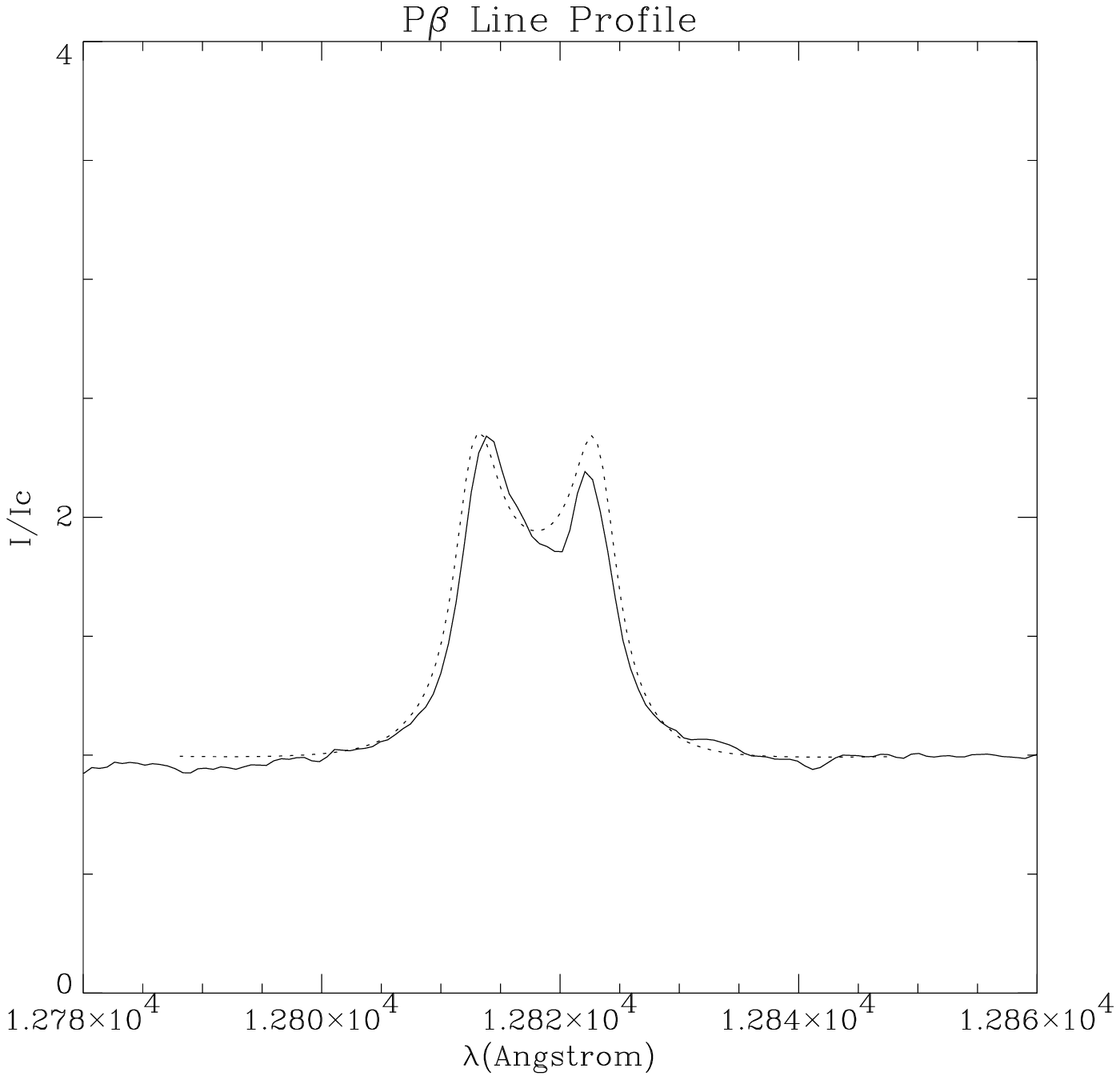}
  \end{center}
  \caption{Left: Observed Pa$\beta$ line profile (mid-2003, solid
  line) compared to the modeled one based on the parameters
  determined by the fit of the 1999 H$\alpha$ and H$\beta$ line
  profiles (dotted line). Right: Observed Pa$\beta$ line profile
  (solid line) and the best fit from the SIMECA code (dotted line). To
  compensate for the effects of non-simultaneous observations, the
  density of the disk was allowed to vary in a small range. This fit has
  been obtained with a disk density lower by 25\% compared to the best
  model based on 1999 H$\alpha$ and H$\beta$ lines. }
\label{pbeta_bad} \label{pbeta}
\end{figure*}

The stellar parameters (Table~\ref{parameters}) are important
mainly to define the distance and the luminosity of the source
(Fig.~\ref{SED}) and represent the first step of the iterative
process leading to a model of the envelope as described below. In
the scope of this model, we have computed the H$\alpha$,
H$\beta$ (Fig.\ref{halpha}) and Pa$\beta$ (Fig.\ref{pbeta}) line profiles.

The critical constraint of this study is based on the fact that
the H$\alpha$ and H$\beta$ lines (Fig.\ref{halpha}), as well as
the the Pa$\beta$ line (Fig.~\ref{pbeta}) are strongly in
emission, whereas from Fig.~\ref{visi_SIMECA} the visibilities are
clearly around unity, i.e. the envelope in the N band is nearly
unresolved. Hence, the circumstellar density must be large enough
to produce Balmer lines in emission while at the same time this
density must be low enough and/or the circumstellar environment
must be well delimited for keeping the envelope unresolved. The
emissive lines of many Be stars like $\alpha$\,Ara exhibit a
characteristic double peaked shape and the 'violet' and 'red'
peaks are denoted V and R. This shape provide stringent constrains
on the wind parameters and particularly on the inclination of the
system.

In order to obtain matching parameters derived from both line profiles
and interferometric measurements we have adopted the following strategy:

\begin{enumerate}
\item The density $\rho_{0}$ at the base of the wind is fixed, having
the greatest influence on the emission line intensity.
\item A range of inclination angles is chosen in order to match the
V and R peak separation, V and R peak intensities and line wings
with the observed profile. In this case, the range is
40-50$\degr$.
\item In the next step, the m1 and m2 parameters are varied, where
$m1$ describes the variation of the mass flux from the pole to the
equator according to:

\begin{equation}
\label{flux}
\phi (\theta )=\phi _{\rm pole}+[(\phi _{\rm eq.}-\phi _{\rm
pole})\sin^{m1}(\theta )]
\end{equation}

where $\theta$ is the stellar colatitude.  The parameter $m2$
describes the variation of the terminal velocity $v_\infty$ as a
function of the stellar latitude:

\begin{equation}
v_\infty(\theta)=V_\infty(\rm pole)+[v_\infty(\rm eq.)-v_\infty(\rm
pole)]\sin^{m2} (\theta)
\end{equation}

where  $V_\infty(pole)$  and $V_\infty(eq)$ are respectively the
polar and equatorial terminal velocities.

\item The ratio between the equatorial and polar mass flux:

\begin{equation}
C1= \frac{\Phi_{eq}}{\Phi_{pole}}.
\end{equation}

\noindent is typically between $10^1$ and $10^4$ (Lamers and Waters
\citealp{lamers}), having a large influence on the wings of the line
profile.

\item Following the above procedure, the H$\alpha$ and H$\beta$ line
profiles are fitted simultaneously to obtain the general envelope
parameters.

\item With the parameters reproducing well the 1999 appearance of
  H$\alpha$ and H$\beta$, the modelled Pa$\beta$ line profile has a
  larger intensity than actually observed in 2003
  (Fig.~\ref{pbeta_bad}). In the following, we assumed that the only
  parameter that might have changed between 1999 and 2003 is the
  density at the base of the envelope. In fact, with a base density
  decreased by about 25~\%, a better agreement between the observed and
  modeled Pa$\beta$ profiles is obtained (Fig.~\ref{pbeta}).

\item The comparison between the theoretical SED we obtained and the
MIDI and ISO data (Fig.~\ref{SED}) is used to constrain the
distance of $\alpha$~Ara. The adopted a distance is based on the
hypothesis on the star parameters and the fitted parameters of the
circumstellar environment of $\alpha$ Ara which settle the
infrared excess and the visual reddening (see discussion
Sec.~\ref{sec:dist}).

\item In a final step, we compute the intensity maps in spectral
  channels of 0.2~$\mu$m. Using the distance estimated in the previous
  step, we obtain theoretical visibilities which are compared to the
  VLTI/MIDI data.  We found that the critical parameters to fit the
  observed visibilities are the distance to the star, the inclination
  angle, the $m1$ value and the density at the base of the wind
  $\rho_{0}$.
\end{enumerate}

\noindent During the above described procedure, more than 500
models were computed using the SIVAM II Alpha ES45 EV68
workstation, build on 3 servers with each four 1 GHz CPU and 24 GB
of RAM at the Observatoire de la C\^ote d'Azur. To compute one
model it takes about 20~min of CPU time.\\


\begin{figure}
  \begin{center}
      \includegraphics[height=7.cm]{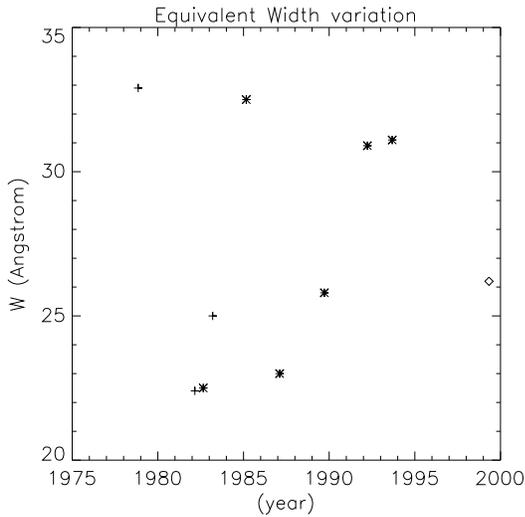}
  \end{center}
  \caption{Variation of the H$\alpha$ equivalent width (EW) with
  time. The data were taken from Dachs et al. \cite{dachs}: crosses;
  Hanuschik et al. \cite{hanuschik}: stars and this work: diamond.}
\label{varia}
\end{figure}

The variability of the circumstellar environment of $\alpha$Ara is
traced by the variations of the H$\alpha$ line equivalent width
(EW) reported in the literature by Dachs et al. \cite{dachs};
Hanuschik et al. \cite{hanuschik} and this work. From
Fig.~\ref{varia} it is evident that we cannot straightforwardly
use parameters derived from the 1999 H$\alpha$ and H$\beta$ line
profiles to model the Pa$\beta$ line profile observed in 2003,
i.e. close to our interferometric run. It is most likely that some
physical parameters have changed between these two epochs.

\section{Results and Discussion}
\label{secresult}
\subsection{Theoretical line profiles and SED}

\begin{figure}
  \begin{center}
      \includegraphics[height=7.cm]{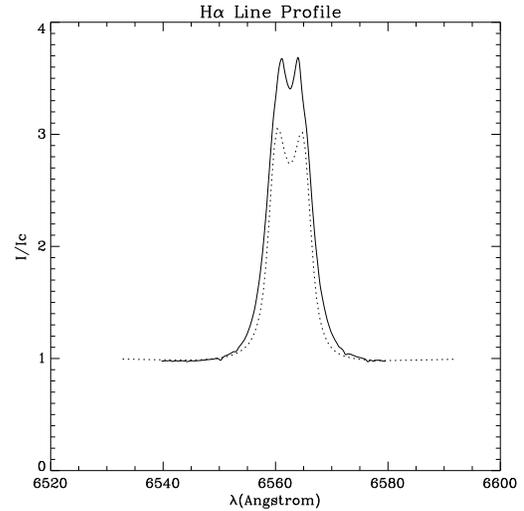}
  \end{center}
  \caption{H$\alpha$ line profile obtained in April 1999 with the {\sc
  HEROS} instrument at la Silla (Chile) and the SIMECA fit (dotted
  line) for the base density used to fit the 2003 Pa$\beta$ profile.
  The strength of the line is comparable of the one reported in 1982
  or 1987, for instance see Fig.~2 by Hanuschik et al. (\citealp{hanuschik})}
\label{halpha_fitpb}
\end{figure}

The H$\alpha$ line observed 1999 shows strong emission of $I/I_{\rm c}
\approx 3.7$, which is well reproduced by SIMECA with the stellar
parameters given in Table~\ref{parameters} and the envelope parameters from
Table~\ref{results}  (Fig.~\ref{halpha}).

With a wind base density decreased by 25\,\%, producing a good
agreement for the Pa$\beta$ line profile (Fig.~\ref{pbeta}), the
model gives an H$\alpha$ line profile with a I/I$_{c}$ ratio
$\approx$3 (Fig.~\ref{halpha_fitpb}). Unfortunately, we do not
have H$\alpha$ line profiles taken contemporary to the VLTI
observations that could confirm such decrease.\\

Even though Fig.~\ref{pbeta} shows that the Pa$\beta$ line profile
modelled with SIMECA is in good agreement with the observed one, the
modelled $V/R$ is less than unity, whereas the observed one is $V/R
\approx 1.17$. SIMECA is based on a radiative wind model for which the
gas is outflowing from the star with a polar and equatorial velocity
of respectively 2000 and 170 km\,s$^{-1}$. In order to obtain a
$V/R>1$, it would be necessary to introduce an asymmetry like a global
one-armed oscillation, as seen by Berio et al. \cite{berio} and Vakili
et al \cite{Vakili98} for the $\gamma$~Cas and $\zeta$~Tau equatorial
disks.  Nevertheless, the global agreement of observation and model
both in profile shape and intensity indicates that, under the general
assumptions of the SIMECA model, the global kinematics and density
distribution within the envelope are well reproduced by the assumed
model parameters.

The computed SED between 2 and 10~$\mu$m is shown in Fig.~\ref{SED},
together with the data recorded by VLTI/MIDI (8-13.5~$\mu$m) and the
ISO satellite (2.5-11~$\mu$m).  The observed SED and the modelled
curve match for a source distance of 105~pc. This distance is
significantly larger than the one obtained from the Hipparcos
satellite (see discussion in Sect.~\ref{sec:discussion}).  In
Fig.~\ref{SED} the continuum emitted by the central star is
approximated as a $T_{\rm eff}$=18\,000~K black body radiation
plotted as a dashed line. The total, i.e. free-free and free-bound
emission from envelope + star, is indicated by the dotted line. Clearly,
the infrared excess produced by the circumstellar envelope is mandatory
in order to fit the observed data (solid line).

\subsection{Theoretical visibilities}

\begin{figure*}
  \begin{center}
      \includegraphics[height=7.cm]{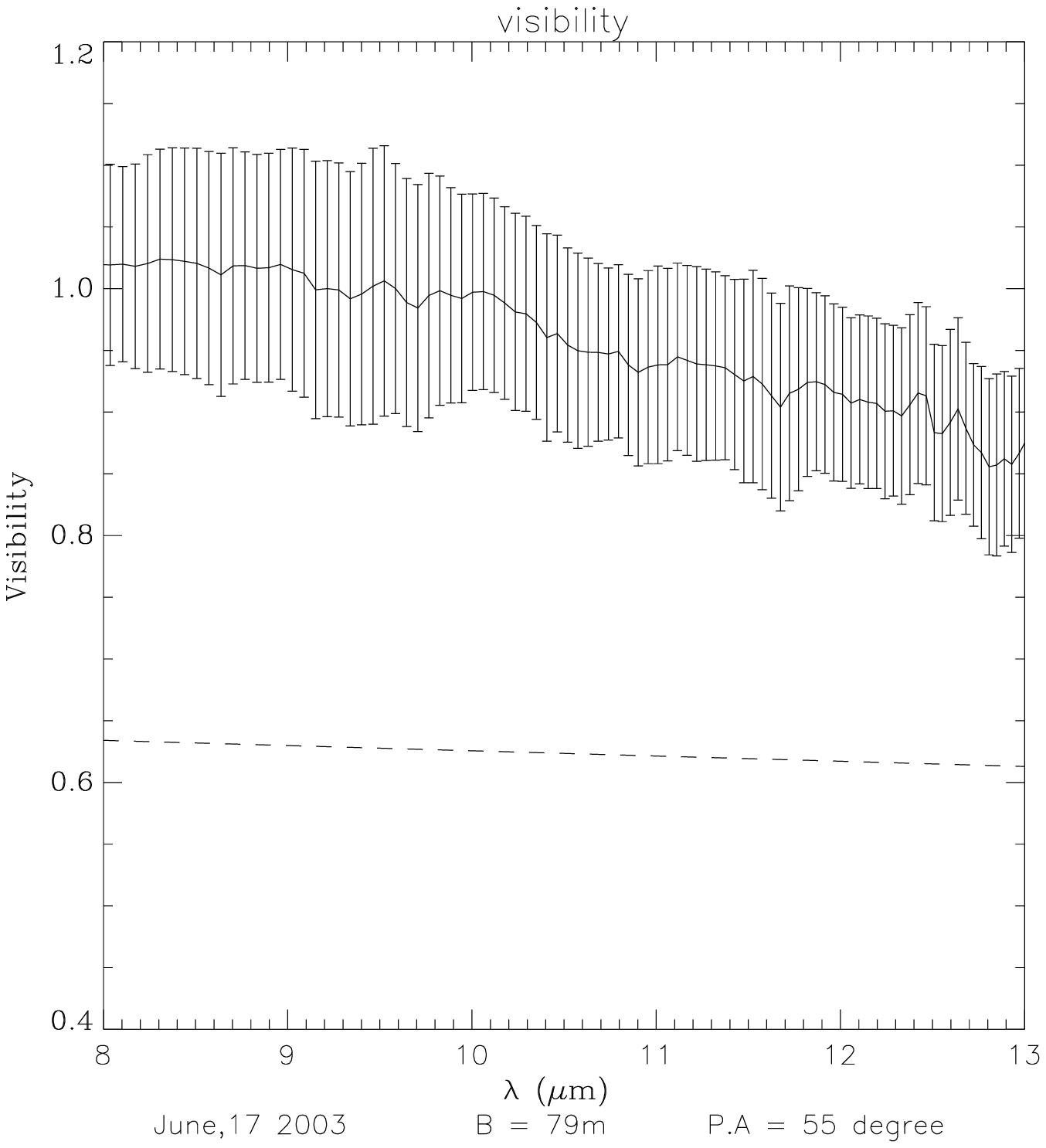}
      \includegraphics[height=7.cm]{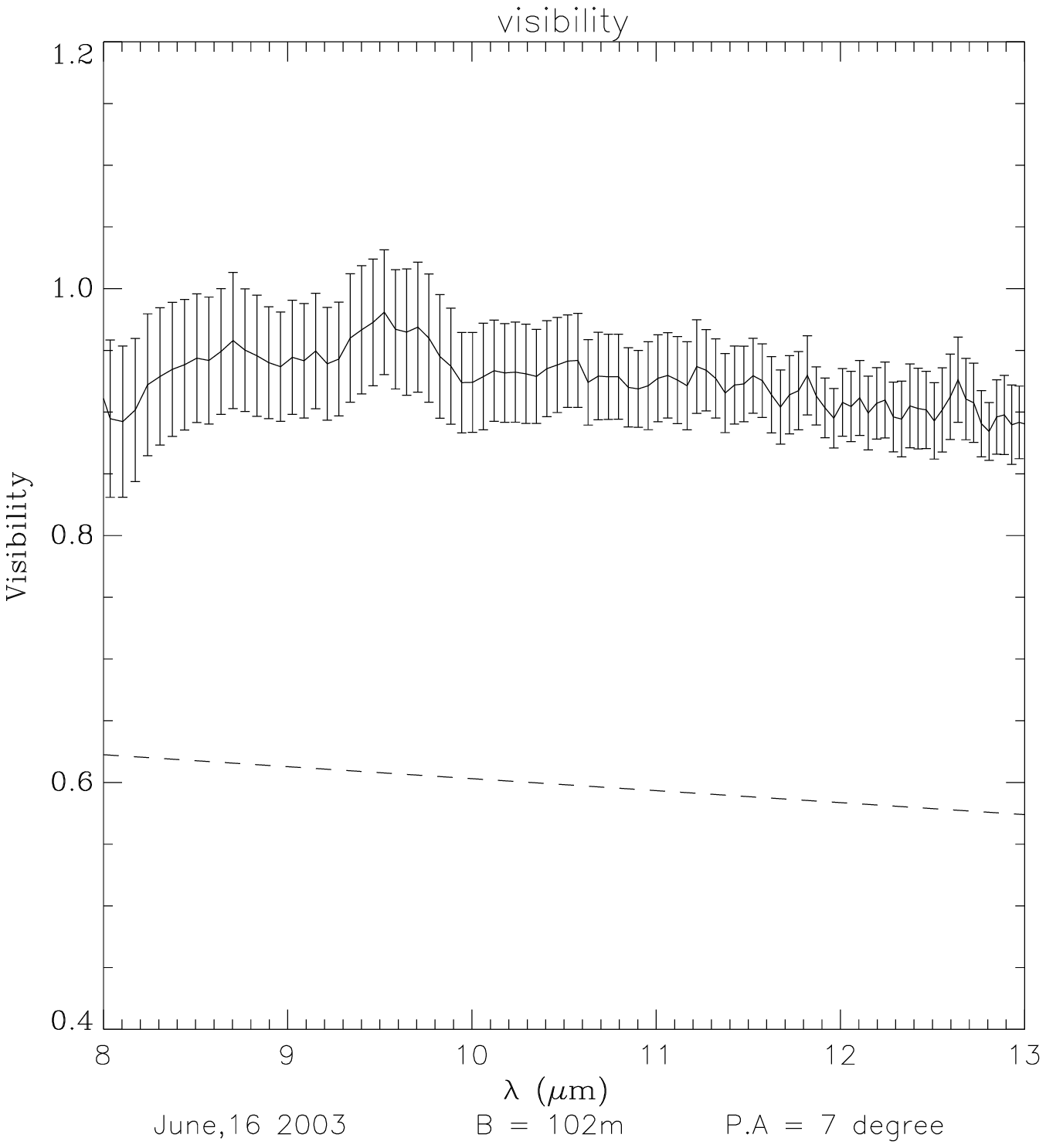}
  \end{center}
  \caption{VLTI/MIDI data for the 79~m (left) and 102~m (right) baselines and theoretical visibilities from SIMECA (dashed line)
  for the distance of 105 pc estimated from our SED fit (see
  Fig.~\ref{SED}). The error bars are equivalent to one sigma.}
 \label{visi_SIMECA}
\end{figure*}

The parameters obtained by the above modeling of the emission line
profiles, and the density required to fit the 2003 Pa$\beta$ line,
can now be used to compute the expected visibility curves for a
distance of 105 pc, as estimated from the fit of the SED.  The
resulting visibility curves are plotted in Fig.~\ref{visi_SIMECA},
clearly showing that the modeled envelope should be well resolved
at 79~m and 102~m (mean visibility $V_{mean}$ $\approx$ 0.63)
whereas the VLTI/MIDI data without any doubt have hardly resolved
the target ($V_{mean}$ $\approx$ 1). However, the observed
visibility curves in Fig.~\ref{visi_SIMECA} may indicate that the
envelope is more resolved for the longer wavelengths. This effect,
if true, is more or less reproduced by our model and is more
obvious for the 102~m baseline where the theoretical visibility is
0.62 at 8~$\mu$m and 0.57 at 13~$\mu$m.

\begin{figure*}
  \begin{center}
      \includegraphics[height=7.cm]{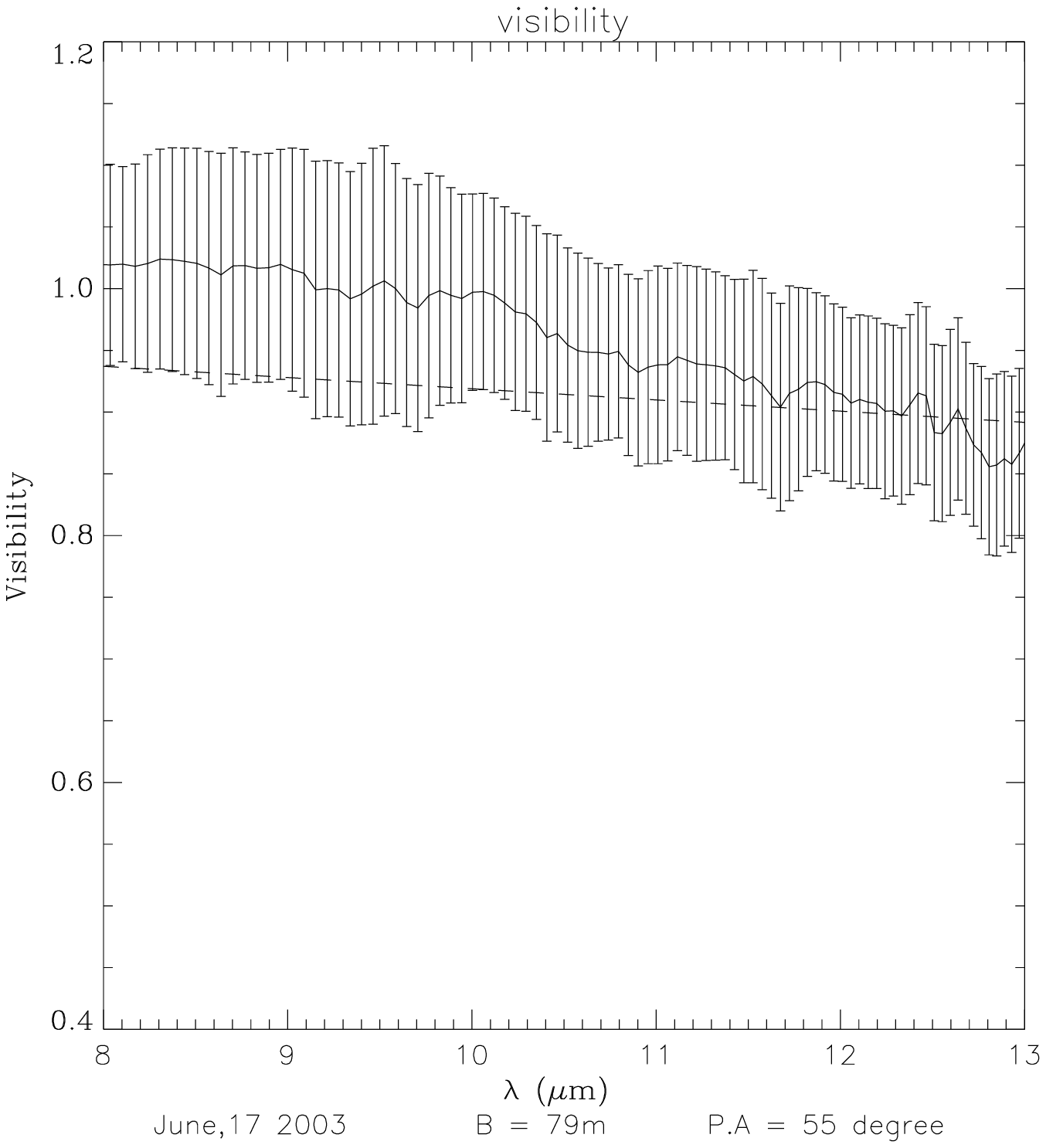}
      \includegraphics[height=7.cm]{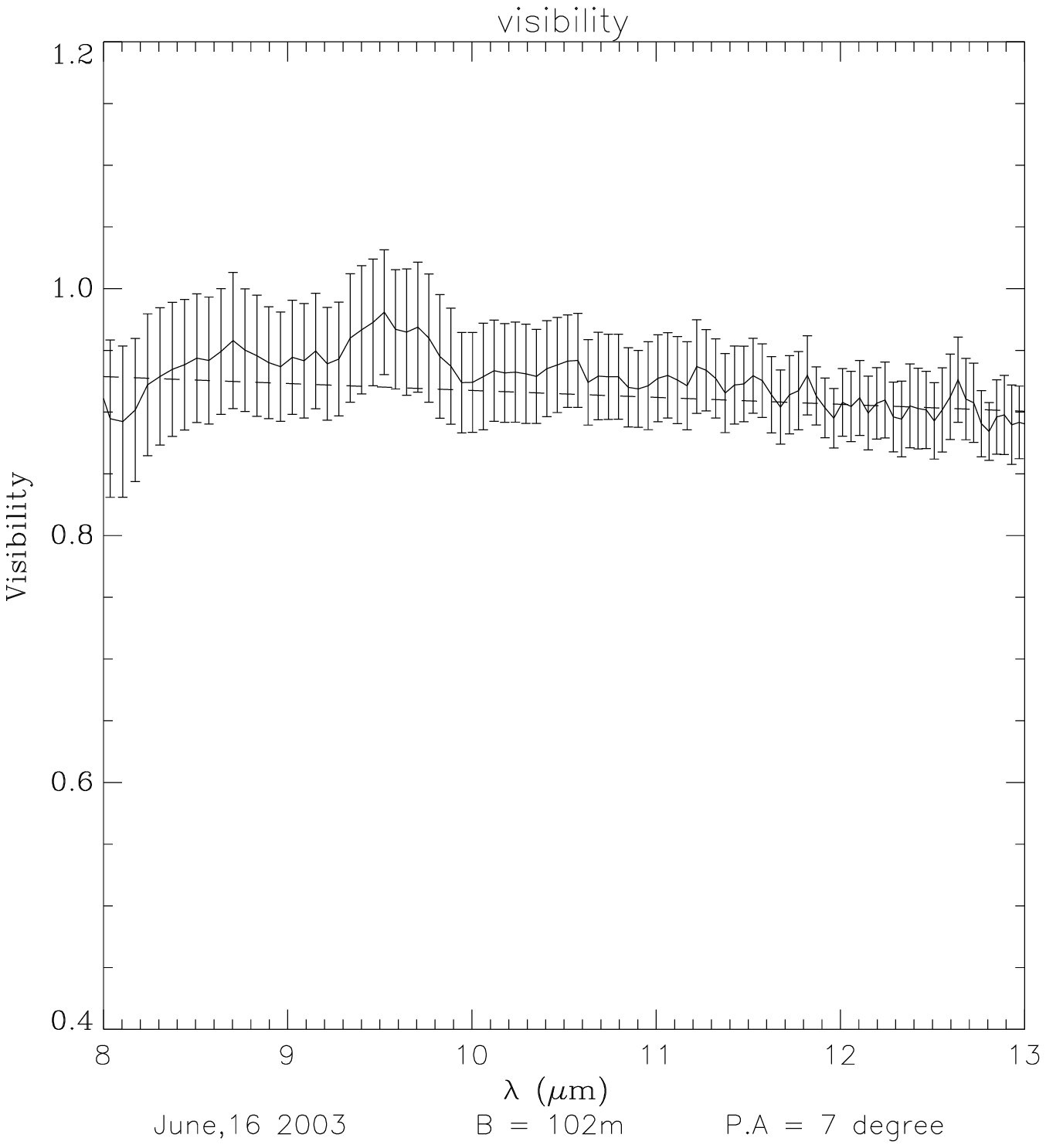}
  \end{center}
  \caption{VLTI/MIDI data for the 79~m (left) and 102~m (right) baselines and theoretical
  visibilities from SIMECA (dashed line) for a truncated disk at 22
  R$_{\star}$ by a possible companion. The error bars are equivalent
  to one sigma.}
 \label{visi_TRUNC}
\end{figure*}

Nevertheless, we did not succeed to obtain model parameters
reproducing both the Balmer lines and the visibility curves at the
same time: At a distance of 105 pc the model gives an envelope
which is clearly resolved, if the constraint that the Pa$\beta$
line is in emission with I/I$_{c}$ $\approx$ 2.3 is met. One
possibility would be to increase the distance of the star up to
122 pc, a distance suggested by Cohen et al. \cite{cohen}.
However, we consider this value as too high by comparison to the
Hipparcos distance (see also the discussion in
Sec.\ref{sec:discussion}).

The modeled envelope is more or less spherical since the input
parameter $m1=0.3$ corresponds to an opening angle of 160$\degr$
(see Stee \citealp{Stee4}), producing only a negligible visibility
difference for the envelope seen parallel or perpendicular to the
polar axis. Nevertheless, this small flattening may produce the
small intrinsic linear polarization of Pl$\approx$0.6\% measured
by McLean \& Clarke \cite{mclean} and Yudin et al. \cite{yudin},
who found a polarization angle PA=172$^\circ$. The inner disk
orientation is, therefore, expected to be perpendicular to this
direction at 82$^\circ$. This means that the longer baseline was
unfortunately oriented at about 70$^\circ$ from the expected major
axis of the disk. The diagram of the $\alpha$~Ara circumstellar
envelope projected onto the sky plane (with an arbitrary
oblateness) and the VLTI baseline positions is given in
Fig.~\ref{schema}.

\begin{figure*}
  \begin{center}
      \includegraphics[height=7.cm]{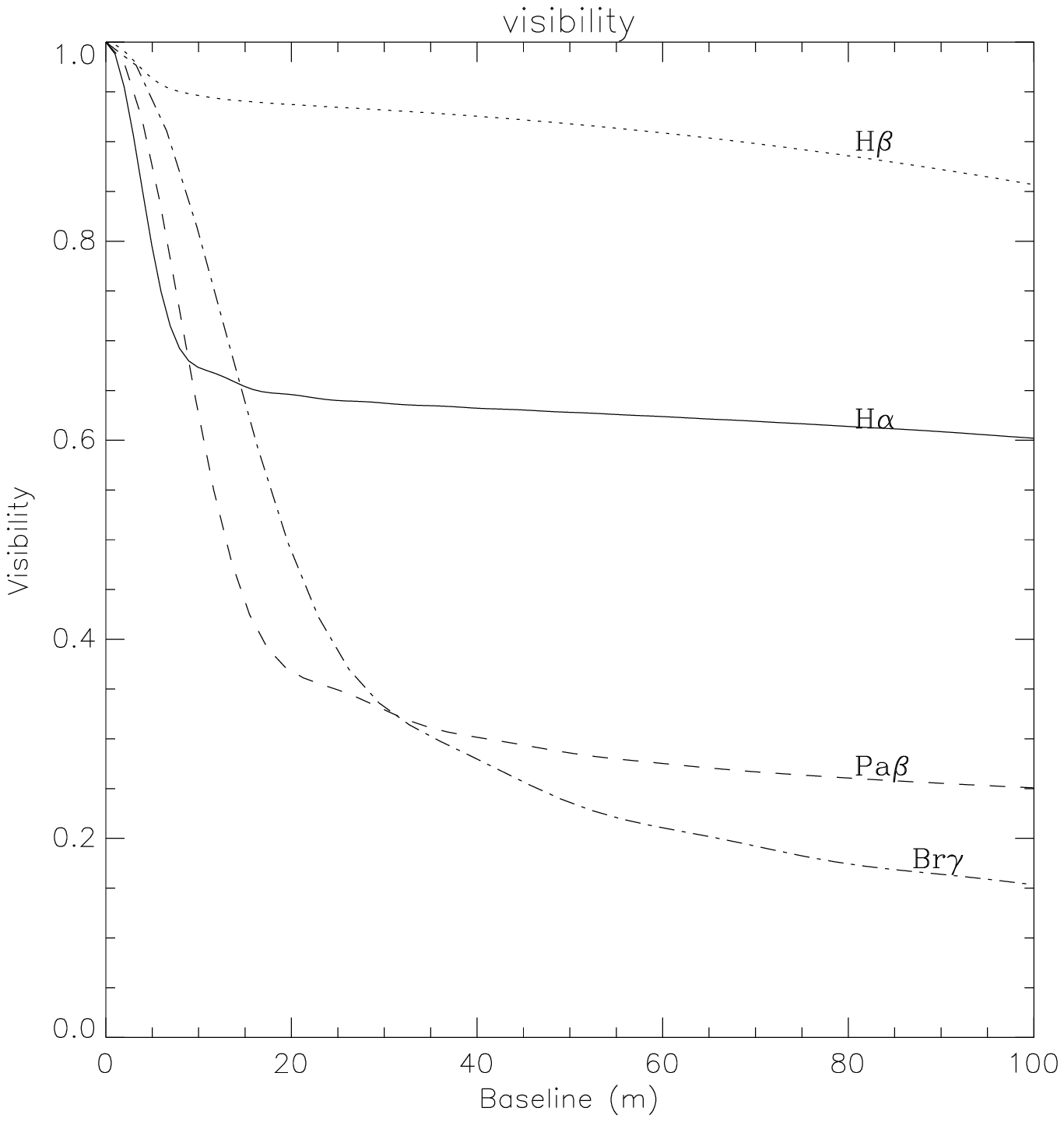}
      \includegraphics[height=7.cm]{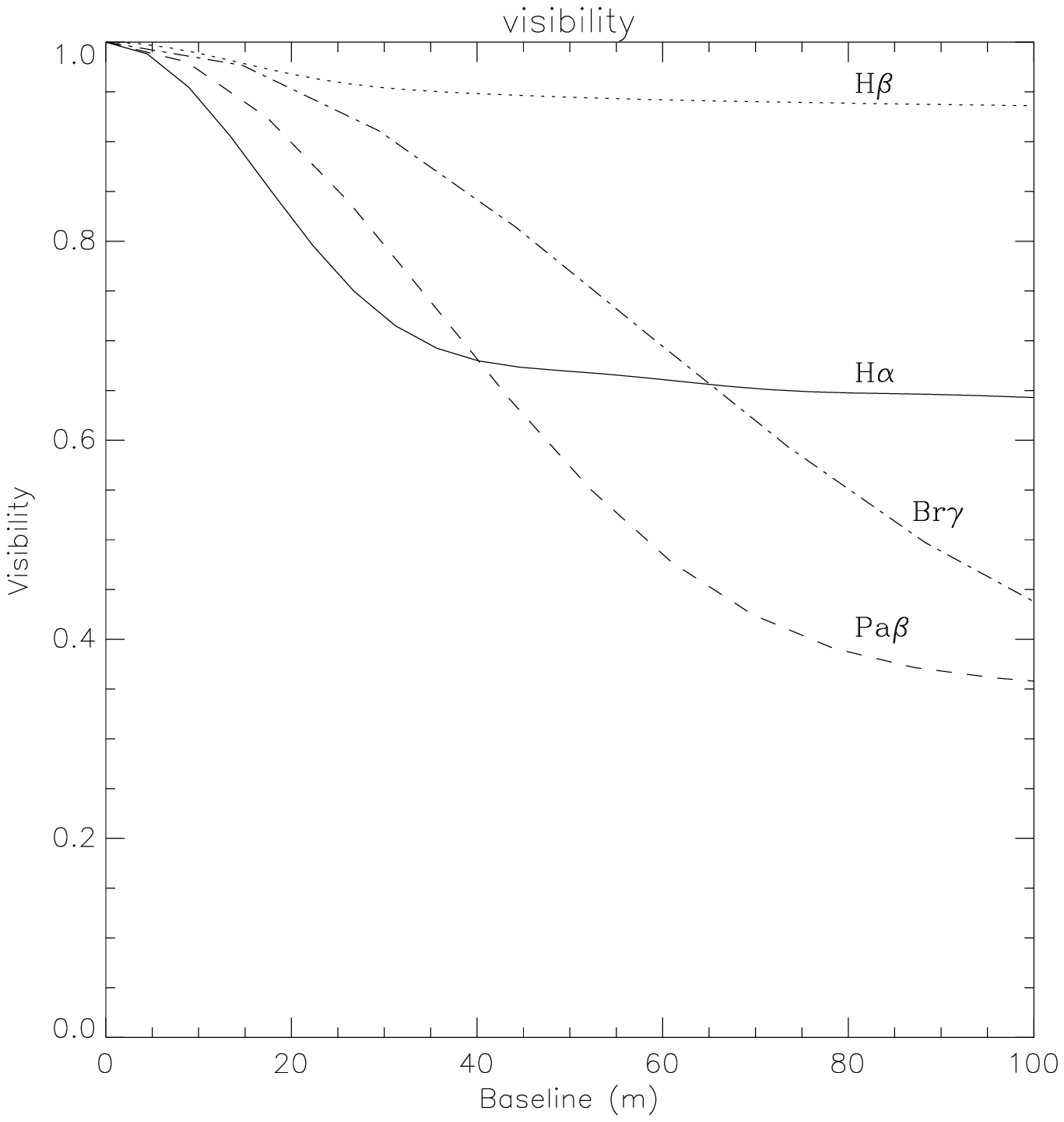}
  \end{center}
  \caption{Theoretical visibilities for the H$\alpha$, H$\beta$,
  Pa$\beta$ and Br$\gamma$ lines for baseline orientations
  corresponding to the ones observed on June 16, i.e.\ PA= 7$\degr$
  onto the sky plane, for both scenarios discussed. On the left
  side we show model data for a disk with the parameters which fits
  the 2003 Pa$\beta$ line profile, but not in agreement with the
  VLTI/MIDI visibilities (Fig.~\ref{visi_SIMECA}).  On the right side
  model data for a disk truncated at 22~R$_{\star}$, e.g.\ by a
  putative companion, is presented, being in agreement with the
  observed visibilities (Fig.~\ref{visi_TRUNC}).}
 \label{visilines}
\end{figure*}

\begin{figure}
  \begin{center}
      \includegraphics[height=7.cm]{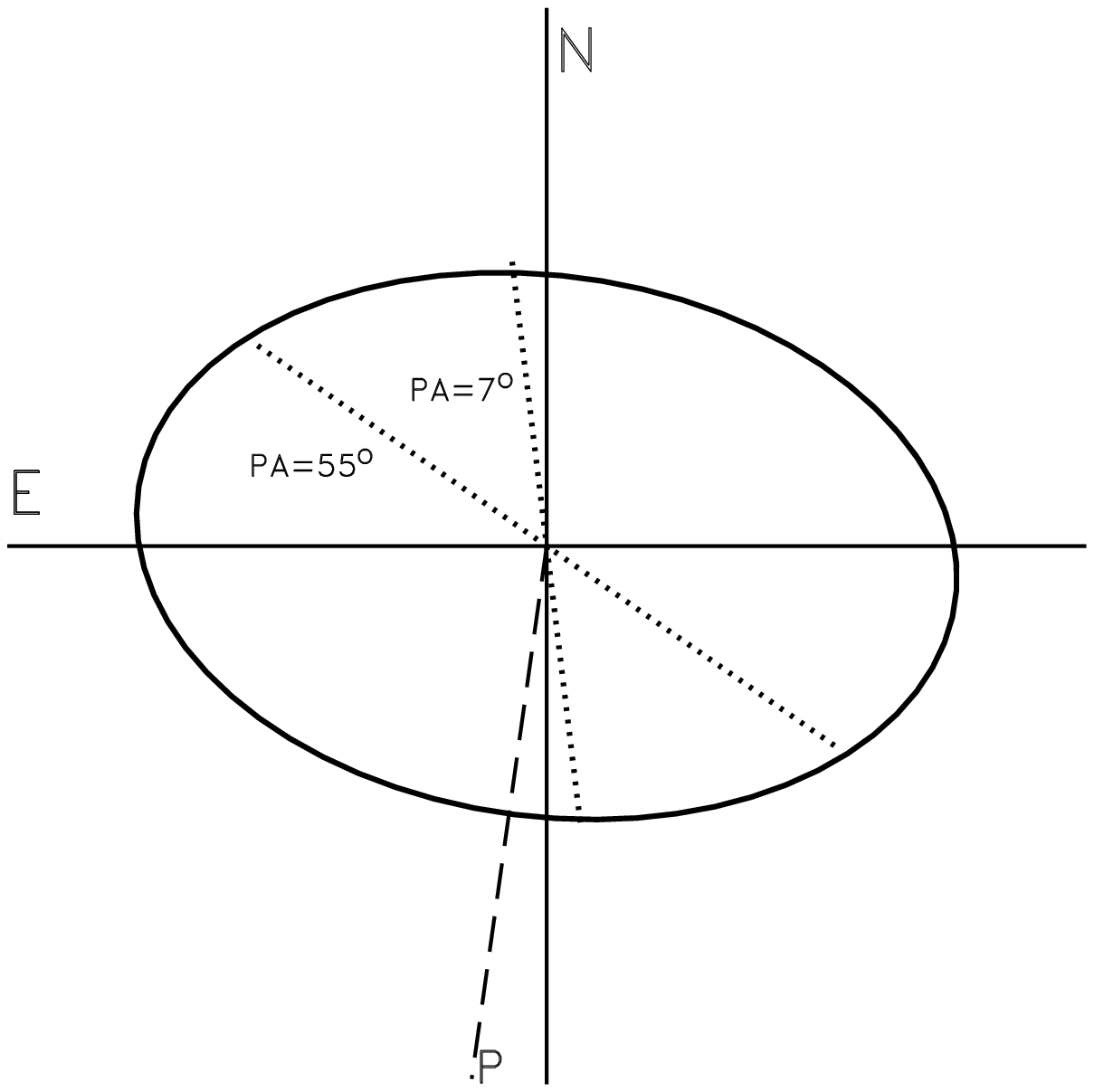}
  \end{center}
  \caption{Diagram of $\alpha$ Ara circumstellar envelope projected onto the
  sky plane (with arbitrary oblateness) based on polarization (P) measurements
  (PA=172$\degr$) obtained by McLean \& Clarke \cite{mclean} and Yudin et
  al. \cite{yudin}. The actually observed VLTI baseline positions are shown for
  June, 16: PA = 7$\degr$, B=102m and June, 17: PA = 55$\degr$, B=79m
  following the polarization (P) measurements (PA=172$\degr$) obtained by
  McLean \& Clarke \cite{mclean} and Yudin et al. \cite{yudin}. }
 \label{schema}
\end{figure}

Considering these difficulties, we have computed a model with a truncated disk. The
radius where the truncation occurs was set to 22~R$_{\star}$, which was
derived using a distance of 105~pc and the fit of the Pa$\beta$ line.
(Fig.~\ref{visi_TRUNC}).
The theoretical visibilities in the H$\alpha$, H$\beta$, Pa$\beta$ and
Br$\gamma$ lines are plotted in Fig.~\ref{visilines} for the two
scenarios. For both models it is clear that at shorter wavelengths, with a
102~m baseline, the envelope of $\alpha$~Ara would be well resolved in the
Pa$\beta$ and Br$\gamma$ lines ($V \approx$ 0.2-0.3) even for a truncated disk
($V \approx$ 0.35-0.45). In H$\alpha$ and H$\beta$, $V$ is 0.6 and 0.85,
respectively, for the full disk and $V$ is 0.65 and 0.94 for the truncated
one. At these wavelengths, the emission comes from the inner part of the disk,
which remains unresolved at a distance of 105~pc. Consequently, such a
truncation of the disk would have only a little impact on the H$\alpha$ and
H$\beta$ emitting regions, whereas the IR emission, originating from
more extended regions, is more affected.


The fact that $\alpha$~Ara is nearly unresolved at 8~$\mu$m with a
102~m baseline gives an upper limit for the diameter of the
emitting envelope of $\phi_{max}$= 4$\pm$1.5~mas, i.e.\
14~R$_{\star}$ (for $d=$74~pc) or 20 R$_{\star}$ (for $d=105$~pc)
assuming an uniform disk distribution for the star+disk brightness
distribution. If a gaussian distribution is assumed, the best fit
is obtained with a FWHM=2.5mas although the quality if the fit is
poorer than for the uniform disk hypothesis. Such a radial
extension is in good agreement with the one obtained by Waters et
al. \cite{waters87} from IRAS Far-IR data, with $\phi_{disk}$
$\sim$ 7 R$_{\star}$. Indeed, the hypothesis that the
star+envelope flux distribution is unchanged from 8$\mu$m to
13.5$\mu$m is a crude approximation and these estimated extensions
must be taken as indicative only. It must be stressed out that the
slopes of the visibility curves provided by the model are in good
agreement with the data showing that the flux is more concentrated
at 8$\mu$m than at 13$\mu$m. Moreover, the angular diameter
estimations are mostly based on the visibilities taken with the
longest projected baseline at PA=7$^\circ$ to the East, i.e. more
or less along the predicted polar direction (see
Fig.~\ref{schema}).

In Table \ref{results} we present the results based on the best fit of
the $\alpha$ Ara August 2003 Pa$\beta$ line profile and visibilities in the N
band.

{
\begin{table*}
{\centering \begin{tabular}{l|c|c} \hline
parameter/result    & value & uncertainty\\
\hline Distance & 105\,pc & 10\,pc\\
Inclination angle i & 45\ensuremath{} & 5\\
Photospheric density ($\rho_{phot}$)&1.2 10\( ^{-12} \)g cm\( ^{-3} \) & linked to C1\\
Photospheric expansion velocity& 0.07 km s\( ^{-1} \) & 0.01 \\
Equatorial rotation velocity & 300 km s\( ^{-1} \) & 20 \\
Equatorial terminal velocity & 170 km s\( ^{-1} \) & 20 \\
Polar terminal velocity & 2000 km s\( ^{-1} \) & 500 \\
Polar mass flux & 1.7 10\( ^{-9} \)M\( _{\sun } \) year\( ^{-1} \) sr\( ^{-1} \) & 0.5 10\( ^{-9} \)\\
m1 & 0.3 & 0.1\\
m2 & 0.45 & 0.05\\
C1 & 30 & linked to $\rho_{phot}$\\
Mass of the disk & 2.3 10\( ^{-10} \)M\( _{\sun } \) & - \\
Mass loss & 6.0 10\( ^{-7} \)M\( _{\sun } \) year\( ^{-1} \) & -\\
\hline
\end{tabular}\par}

\caption{Best model parameters for the $\alpha$ Ara circumstellar environment
obtained by fitting the visibilities and the 2003 Pa$\beta$ line profile
with SIMECA\label{results}}
\end{table*}
\par}
\vspace{0.3cm}

\section{Discussion}
\label{sec:discussion}
The characteristics of the disk around $\alpha$~Ara seem to have
been fundamentally stable over the past century. Only the overall
strength of the emission, as traced e.g.\ by the H$\alpha$ EW
exhibit large variations (see Fig.~\ref{varia}). However, the
compilation of spectroscopic observations and the consecutive
study of this disk by MIDI have revealed a number of interesting
and partly unsuspected characteristics.

\subsection{Variability and suspected binarity}
The Hydrogen recombination line profiles of $\alpha$~Ara exhibits
(quasi?)-periodic short-term $V/R$ variations with a timescale of 45 days
(Mennickent \& Vogt, 1991). The timescale for these variations measured in the
HEROS dataset, however, is better accounted for with a larger period of
70-90~days (see below). Superimposed to these variations are changes of the
emission line equivalent widths. For instance, the strength of the H$\alpha$
emission varies from about 3.2 to more than 4.5 in units of the local
continuum in only a few weeks. The period of variations could be the same as
the $V/R$ period, but this cannot be shown from our data and it also has not
yet been noticed by other investigators. The authors studying the $V/R$
variations concentrated most of their efforts on the shape of the line profile
rather than on the absolute strength.  Nevertheless, some interesting cases of
rapid $V/R$ changes, correlated with strong variations of Balmer line
equivalent widths are reported by Mennickent (1989; 1991).

Long-term $V/R$ variations are common in Be stars, being attributed to a
one-armed oscillation in the Keplerian disk. Rapid $V/R$ variations,
however, have not been extensively discussed in the literature. Mennickent
(1991) associates fast quasi-periodic variations (P$\leq$100d) to the smallest
disks, i.e.\ the disks with the lowest equivalent widths in H$\alpha$. Okazaki
(1997) found that for early-type Be stars around B0 radiative effects can
explain the confinement of this one-armed oscillations with typical periods of
about 10~yr. Nevertheless, this model would have difficulties to explain more rapid $V/R$
variations. Another possibility  for such variable asymmetry is the
presence of a companion, orbiting at distances larger than the envelope
extension. Then, these quasi-periodic line profile variations should be somehow
related to the orbital parameters of the putative secondary.

Nevertheless, with respect to the photospheric absorption lines $\alpha$\,Ara is unusually
stable among earlier type Be stars. In a sample of 27 early Be stars observed
with {\sc Heros} observations, only two stars, including $\alpha$\,Ara, did
not show line any detectable profile variability (Rivinius et al. 2003).

\begin{figure}[tb]
\includegraphics[angle=270,width=8.8cm,clip]{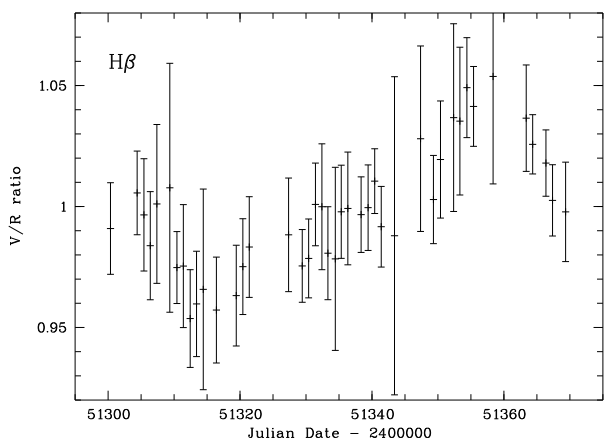}
\caption[+]{\label{fig:Hbet_VtoR}Variation of the height ratio of
violet and red emission peaks of the H$\beta$ line during the {\sc
Heros} observations 1999}
\end{figure}

\begin{figure}[tb]
\includegraphics[angle=270,width=8.8cm,clip]{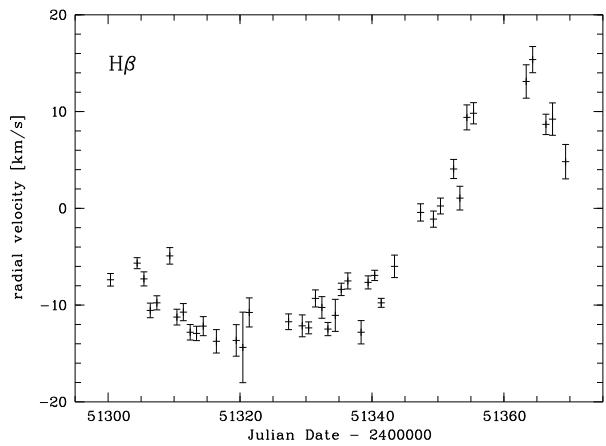}
\caption[+]{\label{fig:Hbet_RV}Radial velocity changes of the
central depression of H$\beta$ during the {\sc Heros} observations
1999}
\end{figure}

However, this is quite different for the circumstellar emission lines. Next to slow,
possibly cyclic long-term changes typical for Be stars, there are changes on a
shorter timescale. In particular, the peak-height-ratio of the violet and red
peaks of the higher Balmer lines have made one full cycle of low amplitude
(about 0.95 to 1.05) during the 69 days of the {\sc Heros} observations (see
Fig.~ \ref{fig:Hbet_VtoR}). At the same time, the radial velocity of the
emission component of the Balmer lines was changing in a cyclic way as
well. For instance, Fig.~\ref{fig:Hbet_RV} shows this effect for the central
absorption of the H$\beta$-emission, but the same is seen if the radial
velocity of the emission peaks is traced, or if other Balmer lines are
investigated. Unfortunately, the available data are not good enough to claim
such a radial velocity change in the weaker emission lines or in the
photospheric absorption lines. The uncertainty of measuring the radial
velocity of such lines is much higher than the amplitude seen in
Fig.~\ref{fig:Hbet_RV}.  Mennickent et al.\ (1991) have observed similar
changes and gave a cycle length of 0.13 years, or 45 days. However, their
Fig.~2 would also support a cycle length of 70 to 80 days, as present in the
{\sc Heros} data of 1999.

Such behavior is known from a few other Be stars as well. In the cases
investigated thoroughly, it was shown to be linked to binarity,
although the mechanisms are not yet known (see for instance Koubsky et
al. 1997). A search for any spectral contribution of such a
hypothetical companion in the phase-binned {\sc Heros} spectra of
$\alpha$\,Ara did not return a positive result.

Nevertheless, since a companion as a potential origin for an outer truncation
of the disk is important in light of the above results, the properties of such
a system are estimated from the following: Using the stellar parameters of
Table~\ref{parameters}, i.e.\ a mass of $10\,{\rm M}_\odot$, a 70\, day period
would give a radius of about $154\,{\rm R}_\odot$, assuming a circular orbit
of a companion with negligible mass. With $R_\star=4.8\,{\rm R}_\odot$,
this corresponds to about 32 stellar radii, which is in agreement with the
estimate based on the MIDI/VLTI data for disk truncated at 25~R$_\star$, i.e.\
somewhat smaller than the companion orbit.

Such an orbital dimension can also be estimated by integrating the
radial velocity amplitude detected in H$\beta$ of the order of
12-16\,km\,s$^{-1}$. Assuming a circular orbit, a period of
69~days and testing two inclination angles of $i=30^\circ$ and
$60^\circ$, the mass of the companion would range from
1.4~M$_{\odot}$ (F2-4V) and 2.8~M$_{\odot}$ (A2-4V). The
corresponding Roche lobe radii R$_{\rm Roche}$ are roughly between
15 and 20~R$_\star$ although we point out that the radial velocity
curve shown in Fig.\ref{fig:Hbet_RV} would suggest an eccentricity
of about 0.2-0.3, affecting the previous estimation.

However, the radial velocities derived from emission lines cannot be
taken at face value. Depending on the systemic properties, the radial
velocities measured in the central inversion of the circumstellar
emission follow the actual radial velocity curves of the stars only
loosely.  The above orbit size and period, again for the circular case
and negligible mass of the companion, result in an orbital velocity of
about 18\,km\,s$^{-1}$, which is in the same order of magnitude as the
amplitude measured in the emission lines. In some well-investigated
binaries like 59~Cyg (Harmanec et al. 2002, Rivinius \& \v{S}tefl
2000, P$\sim$29d) or $\phi$~Per (Hummel \& \v{S}tefl, 2001, \v{S}tefl
et al. 2000, P$\sim$127d), the emission of the Balmer lines is indeed
in phase with the orbital motion.

Until now, no evidence has been reported for a possible companion around
$\alpha$~Ara, but it should be kept in mind that it is difficult to search for
low mass companions around bright stars, such as Be stars (see for instance
the case of $\gamma$~Cas, Harmanec et al.  2000; Miroshnichenko, Bjorkman, \&
Krugov 2002, period P$\sim$204d). The X-ray flux from $\alpha$~Ara is not
peculiar and similar to that of normal B stars
(L$_X$=3.75~10$^{28}$erg~s$^{-1}$, assuming d=122pc\footnote{This implies a
luminosity L$_X$=1.02~10$^{29}$erg~s$^{-1}$, assuming the Hipparcos distance
d=74pc, which is still in the range of luminosities encountered for this
spectral type by Cohen et al. \cite{cohen}}, Cohen et al.
\citealp{cohen}). The proposed truncation may also help to explain the
stringent non detection of radio emission from $\alpha$~Ara at a level of
0.1~mJy at 3.5cm and 6.3cm (Steele et al. 1998).

Using the NPOI interferometer, Tycner et al. \cite{tycner} have recently
studied the disk geometry of the Be star $\zeta$~Tau, which is also a
well-investigated spectroscopic binary (P$\sim$133d, K$\sim$10\,km\,s$^{-1}$).
They measured the disk extension quite accurately to be well within the Roche
radius. This suggests also that this disk may be truncated.

\begin{figure}
  \begin{center}
      \includegraphics[angle=270,width=0.5\textwidth]{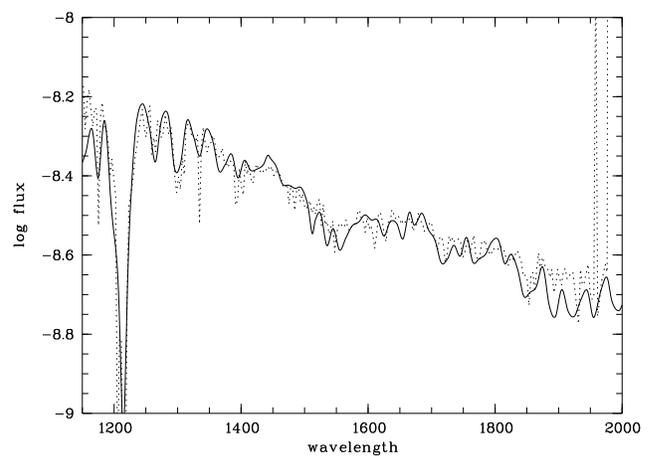}
  \end{center}
  \caption{\label{figIUE}Average of the IUE-spectra of $\alpha$\,Ara
  (dotted) compared to a theoretical spectrum with $T_{\rm
  eff}=18\,000$\,K and $\log g = 4.0$ (solid line).}
\end{figure}

\subsection{Stellar parameters and distance estimation}
\label{sec:dist} The distance used to model the star deviate from
the Hipparcos distance by several sigma. The error bar of the
Hipparcos measurement is 6~pc, and even taking into account a
putative correlated error of the order of 1~mas\footnote{This
correlated error is realistic since Narayanan \& Gould, (1999)
report spatially correlated errors of Hipparcos up to 2~mas in the
direction of the Pleiades and Hyades clusters.}  (Narayanan \&
Gould, 1999, Pan et al. \citealp{pan}), the largest distance
within 1 sigma could be 87~pc.  The distance of 105~pc comes
straightforwardly from chosen stellar parameters, i.e.\ the
absolute magnitude as estimated from effective temperature and
radius. We notice for instance that Fabregat \& Reglero (1990)
obtain a distance of 98~pc from direct $ubvy$ photometry. Zorec \&
Briot (1991) obtain 85~pc, and note that their estimates of Be
star distances seem systematically lower by 20~pc compared to
normal B stars observed in clusters, suggesting that their
distance is a lower limit. Also comparing the flux-calibrated IUE
archival spectra to Kurucz ATLAS9-data gives a distance of about
105~pc (see Fig.~\ref{figIUE}), assuming $R_\star=4.8\,{\rm
R}_\odot$ and neglectable extinction in the UV. The latter is
probably justified since there is no interstellar extinction
towards such a close star (Dachs et al., 1988), and the
circumstellar effects for a non-shell star (see discussion below)
should not affect the UV-regime. We note also that the parameters
of $\alpha$\,Ara and its hydrogen environment provided by the SED
fits of Dachs et al. (1988) are close to the ones presented in
this paper, in particular the estimated circumstellar reddening.

\begin{figure}
  \begin{center}
      \includegraphics[width=0.5  \textwidth]{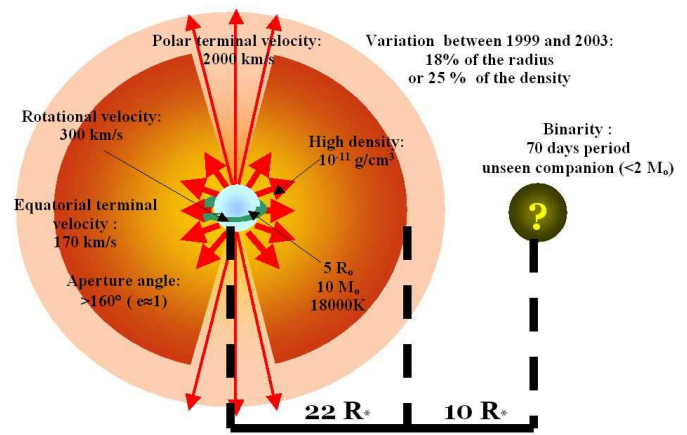}
  \end{center}
  \caption{Schematic view of the $\alpha$ Ara circumstellar environment as
  used in this work (see Table \ref{results}).}
 \label{model}
\end{figure}

The available spectral classifications of the star are all around
B3\,Ve, so that it can be assumed that the effective temperature
is well constrained at about 18\,000\,K. Instead, the main source
of uncertainty in the model may be the radius, which arbitrarily
has been chosen to be 4.8\,${\rm R}_\odot$, based on the
statistical value given by Tokunaga in Allen's ``Astrophysical
Quantities'' (2000).

However, $\alpha$\,Ara has a high $v \sin i$ of about
270\,km\,s$^{-1}$ (Dachs et al. 1990, Yudin 2001, Chauville et al.
2001). With such a large projected rotational velocity, the
hypothesis of a uniform disk radius is questionable and the radius
used is most probably underestimated as demonstrated by the recent
observations of $\alpha$\,Eri (Achernar, Domiciano de Souza et al.
2003). This implies an increase of the illuminating area and
strengthens the difficulty to put the star at the Hipparcos
distance. Without considering any reddening, and keeping the
Hipparcos distance, the radius of $\alpha$\,Ara would be
unrealistically low (below 3.5\,$R_\odot$) or the photosphere
unrealistically cold ($T_{\rm eff}$ of the order of 15\,000\,K).
In Tab.~\ref{tab:compB3} we show selected parameters of the Be
stars $\alpha$\,Ara, $\alpha$\,Eri and the 'normal' (i.e. non Be)
star $\eta$\,UMa which are all classified as B3\,V. Comparing
these values, it becomes evident that, if their Hipparcos
distances are correct, the striking visual flux differences
between these three stars cannot be directly related to a distance
difference only.

\begin{table}[b!]
\begin{center}
\caption{Stellar parameters of three close B3V stars,
$\alpha$~Ara, $\alpha$~Eri and $\eta$~UMa.} \label{tab:compB3}
\begin{tabular}{lccc}

\hline \hline

& $\alpha$~Ara &  $\alpha$~Eri & $\eta$\,UMa\\
\hline
V &  2.9 &  0.5 &1.7\\
Dist.$^{i}$ &  74.3$\pm$6\,pc &  44.1$\pm$1.4\,pc&  30.9$\pm$0.7\,pc\\
$v \sin i$& 270\,km\,s$^{-1}$& 225\,km\,s$^{-1}$ & 150\,km\,s$^{-1}$\\
B-V &  -0.17&-0.17 & -0.17\\
V-K$^{ii}$ & 0.4 & -0.4 & -0.5\\
R$_{\rm eq}$$^{iii}$& 4.8\,R$_\odot$ & 12~$\pm$~0.4\,R$_\odot$& 4.8\,R$_\odot$\\
R$_{\rm pol}$$^{iii}$& 4.8\,R$_\odot$ & 7.7~$\pm$~0.2\,R$_\odot$& 4.8\,R$_\odot$\\

\hline
\end{tabular}
\end{center}\vskip-3mm

{\scriptsize $^{i}$ Hipparcos distance.}\\
{\scriptsize $^{ii}$ From 2MASS photometry .}\\
{\scriptsize $^{iii}$ 4.8 is the statistical value for this a B3V
star, the radius for
$\alpha$~Eri is from Domiciano de Souza et al. 2003.}\\

\end{table}

The reddening of Be stars does not only contain an interstellar term,
which is zero for $\alpha$\,Ara, but also has a circumstellar
contribution. This is a model dependant parameter, explaining the well
known difficulty to calibrate the Be stars distances independently
from Hipparcos measurements.

For instance, Cohen et al.~\cite{cohen} have apparently totally
neglected any circumstellar reddening\footnote{We note also that
their estimated distances for other Be stars like $\delta$\,Cen
seem also systematically overestimated compared to Hipparcos
distances. In contrary their estimated distance of $\alpha$\,Eri
is only 27pc based on a wrong assumption of the radius of this
star (R$_{eq}$=R$_{pol}$=4.1R$_\odot$) which is indeed very
difficult to constrain without any interferometric observations.}
and derived a distance of 122~pc. This difficulty also affects the
present model and an underestimation of the circumstellar
extinction gives a distance too large. On the other hand, the
measured $B-V$ for $\alpha$\,Ara does not indicate a circumstellar
environment dense enough in the line of sight to explain the
difference, since this should be linked to emission line
properties which are not observed.

A large amount of extinction and reddening of the central star
would have to be due to an edge-on disk, and such an inclination
would lead to a large and deep self-absorption in the Balmer
emission lines, the so-called shell appearance. $\alpha$\,Ara,
however, has never been seen in a shell phase. Instead, the
inclination angle is relatively well constrained by the shape of
the Balmer line to a value of about 45$^\circ$.

An inclination of about 45$^\circ$ implies an equatorial
rotational velocity of the order of 380\,km\,s$^{-1}$, since $v
\sin i = 270\,{\rm km\,s}^{-1}$. Using the parameters in
Table~\ref{parameters}, this value is at about 75\% of the
critical velocity. We point out that the equatorial velocity in
the model was considered a free parameter and the best model
provides an equatorial value of 300\,km\,s$^{-1}$, i.e.\ 60\% of
the critical velocity. With such a high rotational velocity, the
von Zeipel effect is becoming important, and thus the continuum
emission must be strongly latitude dependant (see e.g.\ Townsend
et al., 2004, his Fig.\ 3). This may also confuse the line profile
diagnostic and affect the angle determination. Such an effect is
not included in the present model. A value of the inclination
closer to 90$^\circ$, compatible with the large $v \sin i$
observed, may be related to a stronger circumstellar reddening
from the equatorial environment and a lower integrated stellar
flux since the equatorial regions of the star are cooler and
fainter than the polar ones.

Definitely, the Hipparcos distance, if reliable, is a very
constraining parameter for any model of the circumstellar
environment of $\alpha$\,Ara. More interferometric observations
are needed, firstly to constrain the angular size and shape of the
hydrogen emission lines, and secondly to evaluate the angular size
and shape of the underlying star. The forthcoming VLTI/AMBER
instrument will be able to address the first point by studying the
fringe properties in Pa$\beta$ and Br$\gamma$. However, baselines
on the order of 150-200\,m will be mandatory to resolve the
central star, becoming available only with the 1.8m Auxiliary
Telescopes. Their first light is foreseen for 2005.

\section{Conclusion}\label{secconcl}
The first IR interferometric measurements of $\alpha$~Ara suggest
that the size of the circumstellar environment is smaller than
what was predicted by Stee \cite{Stee4}. The fact that
$\alpha$~Ara remains unresolved, but at the same time exhibits a
strong Balmer emission line puts very strong constraints on the
parameters of its circumstellar disk. Independently of the
physical model we used, these measurements put an upper limit of
the envelope size in the N band of $\phi_{\rm max}$= 4 mas, i.e.\
14 R$_{\star}$ if the star is at 74 pc according to the Hipparcos
parallax or 20 R$_{\star}$ if the star is at 105 pc as suggested
by our model, assuming a spherical uniform disk for the
star+envelope brightness distribution. The suspected presence of a
low mass companion could help to understand the limit of the disk
extension as determined by MIDI. A schematic view of the $\alpha$
Ara circumstellar disk, as it was modeled in this work with the
SIMECA-code, is presented Fig.~\ref{model}.



\begin{acknowledgements}
We thank Anne-Marie Hubert for many fruitful discussions and
Jean-Louis Falin for checking the quality of the Hipparcos data.
We also thank Anatoly Miroshnichenko for his helpful comments.
This research has made use of SIMBAD database, operated at CDS,
Strasbourg, France. The paper benefited from the careful reading
and advises from the referee, Markus Wittkowski.
\end{acknowledgements}

\end{document}